\documentclass[reprint,amsmath,amssymb,aps,prb]{revtex4-1}

\usepackage{amsmath}
\usepackage{amssymb}

\usepackage[dvips]{graphicx}
\usepackage{psfrag}

\newcommand{\com}[2]{\left[ #1 , #2 \right]}
\newcommand{\acom}[2]{\left\{ #1 , #2 \right\}}

\newcommand{\ket}[1]{| #1 \rangle}

\newcommand{\bs}{\boldsymbol}
\newcommand{\eps}{\varepsilon}

\newcommand{\vphi}{\varphi}
\newcommand{\be}{\begin{equation}}
\newcommand{\ee}{\end{equation}}
\newcommand{\bea}{\begin{eqnarray}}
\newcommand{\eea}{\end{eqnarray}}

\newcommand{\p}{\prime}

\newcommand{\hc}{\mathrm{h.c.}}
\newcommand{\sgn}{\mathrm{sgn}}

\newcommand{\povertwo}{\frac{\vphi}{2}}

\begin{document}

\title{The Influence of Interference on the Kondo Effect in a Quantum Dot}
\author{Justin Malecki}
\author{Ian Affleck}
\affiliation{University of British Columbia, Vancouver, Canada}

\begin{abstract}
We study the Kondo effect in a model system of a quantum dot embedded in an Aharanov-Bohm ring connected to two leads.  By transforming to the scattering basis of the direct inter-lead tunneling, we are able to describe precisely how the Kondo screening of the dot spin occurs.  We calculate the Kondo temperature and zero-temperature conductance and find that both are influenced by the Aharanov-Bohm ring as well as the electron density in the leads.  We also calculate the form of an additional potential scattering term that arises at low energies due to the breaking of particle-hole symmetry.  Many of our results are supported by numerical analysis using the numerical renormalization group.
\end{abstract}

\maketitle

\section{Introduction}
\label{sec:intro}

A quantum dot in a gated semi-conductor heterostructure separating ballistic leads is known to exhibit some remarkable phenomena.  Most strikingly, at low temperatures, the conductance through the quantum dot increases 
as the temperature is lowered
and can reach the ideal value 
of $2e^2/h$ when the dot contains an odd number of electrons~\cite{glazman1988, goldhabergordon1998, cronenwett1998}. This is 
due to the Kondo effect, involving the screening of the spin 1/2 of 
the quantum dot by conduction electrons in the leads. If 
 there is an additional tunneling path connecting the two leads that does not pass through the quantum dot, then 
some interesting interference phenomena take place~\cite{wiel2000, ji2000, ji2002, kalish2005, zaffalon2008}. 

Previous theoretical work on this problem has studied both extended 
Aharonov-Bohm rings as well as short ``Kondo-Fano'' devices.   The conductance~\cite{bulka2001, hofstetter2001} and thermopower~\cite{kim2002} was found to exhibit an asymmetric Fano-like dependence on the energy level of the quantum dot.  When the quantum dot is tuned to the Kondo regime that favors a local moment, a flux dependent Kondo temperature has been proposed using different methods~\cite{aharony2005, simon2005, yoshii2008} and the high and low-temperature conductance has been described~\cite{hofstetter2001, aharony2005, simon2005, yoshii2008, davidovich1997, gerland2000, konig2002, konik2004}.  While some Numerical Renormalization Group (NRG) 
work was reported, this only studied the electron occupancy of the quantum dot~\cite{hofstetter2001} or the density of states on the quantum dot~\cite{gerland2000}, both of which can be approximately related to the conductance.  It should also be noted that most of these studies assume a particle-hole symmetric dispersion relation and Fermi energy in the leads. 

In this paper, we reexamine the Kondo-Fano device using a combination of 
analytic and NRG methods.  We only consider the Kondo regime where a local moment is favored on the quantum dot. We are able to reproduce many of the published results cited above as well as predicting for the first time non-trivial dependence of the Kondo temperature and zero-temperature conductance on the electron density in the leads.  Such a dependence on electron density has not been investigated before given that a particle-hole symmetric Fermi energy has always been assumed in the leads.  We calculate the generation of additional potential scattering terms that have often been neglected in previous studies but which do lead to small corrections to the zero-temperature conductance.  Numerical confirmation of many of our results is provided for the first time using the NRG.

Our analytic approach begins in \S~\ref{sec:model} with a tight-binding version of the Anderson model together with a 
direct tunneling term between the two leads and factors representing magnetic flux between the two conducting paths. 
Following refs.~\onlinecite{yoshii2008}, \onlinecite{kondo1968}--\onlinecite{cragg1979}, we then perform an exact  
transformation to the ``scattering basis'' which diagonalizes the 
direct tunelling part of the Hamiltonian when the hybridization to the Anderson impurity is turned off.  This gives a Hamiltonian containing 
no direct tunneling term, 
only the hybridization to the impurity,  
 albeit with a more complicated dependence on flux, inter-lead tunneling, and particle momentum. 
The initial Hamiltonian contains two scattering 
channels, the even and odd states, for example. After transforming 
to the scattering basis, only one linear combination of these 
appears in the Anderson hybridization; we refer to it as the 
``screening channel''.  

As we are primarily interested in the Kondo regime of the quantum dot, we perform a Schrieffer-Wolff transformation in the screening channel basis to obtain an effective Kondo model with an additional potential scattering term $K_R$ that is of order the bare Kondo coupling and which vanishes (to this order) when the dot level is tuned to the symmetric value of $\eps_d = -U/2$ (these terms are defined in eq.~(\ref{dotH})).  This latter term is discussed in \S~\ref{sec:kr}.  Both the generated Kondo interaction and the potential scattering depend on the flux $\vphi$, the strength of the direct inter-lead coupling $t^\p$, and the momentum of electrons in the leads.   From the strength of this Kondo interaction we are able to obtain the dependence of the Kondo temperature on these model parameters as discussed in \S~\ref{sec:TK}. 

Next, in \S~\ref{sec:vr}, 
we integrate out high energy states to obtain 
a low energy effective Hamiltonian. In addition to renormalizing 
the Kondo interaction, this also generates a small 
potential scattering term, $V_R$, of second order in the bare 
Kondo coupling.  Hence, $V_R$ contributes to the leading order term in the potential scattering when $\eps_d = -U/2$ though there may be other contributions as we discuss in the text.  Otherwise, it is the $K_R$ term discussed above that provides the leading order contribution to the potential scattering.  

In \S~\ref{sec:conductance}, we calculate the low temperature conductance in terms 
of the effective S-matrix for low energy electrons via the 
Landauer formula. Below the 
Kondo temperature, a phase shift of $\pi /2$ occurs in the screening 
channel. To a good approximation, the low temperature S-matrix 
is simply determined by the unitary transformation to the 
scattering basis and this $\pi /2$ phase shift in the screening channel. 
A small correction to this S-matrix occurs due to the $K_R$ potential scattering term (or $V_R$ in the case that $\eps_d = -U/2$).
While this approach confirms the results of Ref.~\onlinecite{hofstetter2001} 
in the special case of a half-filled band in the leads, we find that changing 
the electron density in the leads has a large effect.   

We confirm some of  these results by NRG calculations presented in \S~\ref{sec:nrg}.  We only consider the 
simplest case in the Kondo regime, $\epsilon_d=-U/2$, symmetric coupling of the left and right leads to the dot, and with a half-filled band.  We begin by completely describing the renormalization group flow of our model, predicting the form of the various fixed points and crossover energy scales which are then confirmed in the NRG.  Most notably, the Kondo temperature is extracted from the 
energy scale of the Wilson chain at which the crossover to the low 
energy strong-coupling fixed point occurs and agrees excellently with that predicted analytically.
The effective S-matrix 
is compared to the low energy excitation spectrum over various parameter ranges.  We obtain quite good agreement through this comparison, including the small corrections 
from $V_R$.

\section{Model \& Analysis}
\label{sec:model}

\begin{figure}
\begin{center}
\psfrag{t}{$t$}
\psfrag{P}{$\Phi$}
\psfrag{tp}{$t^\p$}
\psfrag{tm}{$t_{d-}$}
\psfrag{tpl}{$t_{d+}$}
\psfrag{e}{$\eps_d$}
\includegraphics[width=0.45\textwidth, clip=true]{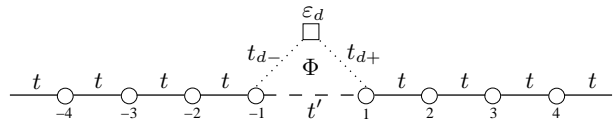}
\end{center}
\caption{\label{fig_lattice} The lattice model described by the Hamiltonian of eq.~(\ref{latticeham}).  }
\end{figure}

We start with a tight-binding model depicted in Fig.~\ref{fig_lattice}.  The Hamiltonian for this model is
\begin{eqnarray}
\label{latticeham}
H & = & H_0 + H_{-+} + H_{td} + H_d \\
\label{latticekinetic}
H_0 & = & - t \left( \left[ \sum_{j = - \infty}^{-2} + \sum_{j=1}^\infty \right]  c_j^\dag c_{j + 1} + \textrm{h.c.} \right) \\
\label{latticedirecttun}
H_{-+} & = & - t^\p \left( c_{-1}^\dag c_1 + \textrm{h.c.} \right) \\
H_{td} & = & - \left[ \left( t_{d-} e^{i \frac{\vphi}{2}} c_{-1}^\dag + t_{d+} e^{-i \frac{\vphi}{2}} c_1^\dag \right) d + \textrm{h.c.} \right] \\
\label{dotH} H_{\mathrm{dot}} & = & \eps_d d^\dag d + U n_{d \uparrow} n_{d \downarrow}.
\end{eqnarray}
Each annihilation operator for the leads, $c_j$, and the Anderson impurity, $d$, is a spinor where the spin indices are implied.  The anti-commutation relationship is $\acom{c_j^\dag}{c_{j^\p}} = \delta_{j j^\p}$.  The number operator for dot electrons of spin $\mu$ is defined as $n_{d \mu} \equiv d_\mu^\dag d_{\mu}$.  The various parameters are described in Fig.~\ref{fig_lattice}.   We will assume that all of the couplings are real.   The magnetic flux has been introduced through the parameter $\varphi = 2 \pi \Phi / \Phi_0$, $\Phi$ being the magnetic flux threading the AB ring and $\Phi_0 = h/e$ being the magnetic flux quantum.    We assume that the magnetic field generating the flux is small enough in the vicinity of the wires so that we may neglect the Zeemen effect in the quantum dot and the leads.

Although such a tight-binding model for the leads is not a very accurate description of leads in a semi-conductor heterostructure on which such geometries are often defined, we use it here as an example of a relatively simple model that contains a natural bandwidth of $4t$.

We now define a basis of even and odd combinations of electron operators
\bea
e_j & \equiv & \frac{1}{\sqrt{2}} \left( c_j + c_{-j} \right), \qquad j > 0 \\
o_j & \equiv & \frac{1}{\sqrt{2}} \left( c_j - c_{-j} \right), \qquad j > 0
\eea
so that the Hamiltonian can be written as
\bea
\label{H0jeo}
H_0 & = & -t \sum_{j=1}^\infty \left( e_j^\dag e_{j+1} + o_j^\dag o_{j+1} + \hc \right) \\
\label{H-+jeo}
H_{-+} & = & - t^\p \left[ e_1^\dag e_1 - o_1^\dag o_1 \right]  \\
\label{Htdjeo}
H_{td} & = & - \frac{1}{\sqrt{2}} \left\{ \left[  t_{de}^\ast e_1^\dag - t_{do}^\ast o_1^\dag \right] d + \hc \right\}
\eea
where we have defined the shorthand notation
\bea
t_{de} & \equiv & t_{d-} e^{-i \frac{\vphi}{2}} + t_{d+} e^{i \frac{\vphi}{2}} \\
t_{do} & \equiv & t_{d-} e^{-i \frac{\vphi}{2}} - t_{d+} e^{i \frac{\vphi}{2}}.
\eea
$H_d$ remains unchanged.  


Immediately we notice that, for the case of zero flux, $\vphi=0$, and symmetric coupling $t_{d-} = t_{d+}$, the model reduces to two decoupled chains, the even channel interacting with the quantum dot and having a potential scattering interaction $-t^\p$ at $j=1$ and the odd channel decoupled from the dot and with a potential scattering interaction $t^\p$ at $j=1$.  However, in the general case of $\vphi \ne 0$, we must analyse both channels together.

If we remove the dot from the system we are left with two independent channels, even and odd, with a potential $\mp t^\p$ at $j=1$.  As shown in Appendix~\ref{sec:phaseshift}, this potential gives rise to two scattering phase shifts $\delta^\pm_k$ in the even/odd channel respectively, the form of which is given at the Fermi surface to be
\be
\label{deltadef}
\tan \delta^\pm \equiv \pm \frac{\tau^\p \sin k_F a}{1 \mp \tau^\p \cos k_F a}
\ee
where $\tau^\p \equiv t^\p / t$ and $\delta^\pm \equiv \delta_{k_F}^\pm$.  Note that, at half-filling when $k_F = \pi / (2a)$, $\delta^+ = - \delta^- = \delta$ where $\tan \delta = \tau^\p$.  These phase shifts will play an important part when we discuss the zero-temperature properties of this system in \S~\ref{sec:physprop}.

Noting that $H_{-+}$ serves as a potential scattering term, we seek to transform to the scattering basis that essentially removes these interactions from the Hamiltonian.  We do this by first introducing the complete set of wavefunctions that solve the Schr\"odinger equation for $H_0$
\be
\phi_j(k) = \sqrt{\frac{2 a}{\pi}} \sin (kja)
\ee
with $a$ being the lattice spacing.  We can expand our operators as 
\bea
e_j & = & \sqrt{\frac{2a}{\pi}} \int_0^{\frac{\pi}{a}} dk \, \sin (kja) e_k \\
o_j & = & \sqrt{\frac{2a}{\pi}}  \int_0^{\frac{\pi}{a}} dk \, \sin (kja) o_k
\eea
so that $\acom{e_k^\dag}{e_{k^\p}} = \acom{o_k^\dag}{o_{k^\p}} = \delta( k - k^\p )$.  The Hamiltonian becomes
\bea
\label{H0eo}
H_0 & = & \int_0^{\frac{\pi}{a}} dk \, \eps_k \left( e_k^\dag e_k + o_k^\dag o_k \right) \\
\label{H+-ab}
H_{-+} & = & \int_0^{\frac{\pi}{a}} dk \, dk^\p \, v_{k k^\prime} \left( e_k^\dag e_{k^\p} - o_k^\dag o_{k^\p} \right) \\
\nonumber
H_{td} & = & - \sqrt{\frac{a}{\pi}} \int_0^{\frac{\pi}{a}} dk \, \sin ka \, \left\{ \left[ t_{de}^\ast e_k^\dag - t_{do}^\ast o_k^\dag \right] d \right. \\
\label{Htdeo}
& & \qquad \qquad \qquad  \qquad \quad \, \, \,  + \hc \bigr\}
\eea
where we have defined 
\bea
\eps_k & \equiv & -2 t \cos ka \\
\label{vkkp}
v_{k k^\p} & \equiv & - \frac{2 t^\p a}{\pi} \sin ka \sin k^\p a.
\eea

With $H_{-+}$ written in such a simple form, we now transform to the scattering basis.  Ignoring $H_{td}$ for the moment, we note that the only difference between the $e$ and $o$ channels is the sign of the $v_{k k^\p}$ interaction.  Hence, we define the scattering basis
\bea
\label{qak}
q_{ek}^\dag & \equiv & \int_0^{\frac{\pi}{a}} dk^\p \phi_{k^\p}^{+(k)} e_{k^\p}^\dag \\
\label{qbk}
q_{ok}^\dag & \equiv & \int_0^{\frac{\pi}{a}} dk^\p \phi_{k^\p}^{-(k)} o_{k^\p}^\dag
\eea
where 
\be 
\label{phikkp}
\phi_{k^\p}^{\pm(k)} \equiv \delta(k - k^\p) + \frac{T^\pm_{k^\p k}}{\eps_k - \eps_{k^\p} + i \eta}
\ee
and $\eta$ is a positive, infinitesimal parameter.  It is shown in detail in Appendix~\ref{sec:tmatrixderive} that $T_{k k^\p}^\pm$ is given by
\bea
\label{Tkkp}
T_{k k^\p}^\pm = \pm \frac{v_{k k^\p}}{1 \pm \tau^\p e^{- i k^\p a}}.
\eea
Thus, the Hamiltonian greatly simplifies in the $q_{ek}$, $q_{ok}$ basis to
\bea
\label{H0qaqb}
H_0 & = & \int_0^{\frac{\pi}{a}} dk \, \eps_k \left( q_{ek}^\dag q_{ek} + q_{ok}^\dag q_{ok} \right) \\
H_{-+} & = & 0 \\
\nonumber
H_{td} & = & - \sqrt{\frac{a}{\pi}} \int_0^{\frac{\pi}{a}} dk \left\{ \left[ t_{de}^\ast {\Gamma_k^+}^\ast q_{ek}^\dag -  t_{do}^\ast {\Gamma_k^-}^\ast q_{ok}^\dag \right] d \right. \\
\label{Htdqaqb}
& & \qquad \qquad \qquad \, \, + \hc \biggr\}
\eea
where
\be
\label{Gammakdef}
\Gamma_k^\pm \equiv \int_0^{\frac{\pi}{a}} dk^\p \sin k^\p a \, \, \phi_{k^\p}^{\pm(k)} =  \frac{\sin ka}{1 \pm \tau^\p e^{-i ka}}.
\ee
The last equality is proven in Appendix~\ref{sec:tmatrixderive}.

With the potential scattering Hamiltonian $H_{-+}$ vanishing due to the transformation to the scattering basis, we are now free to rotate the basis once more to the channel that couples directly to the impurity and its orthogonal complement.  In this way, anticipating our discussion on the Kondo effect, we define the screening channel
\be
\label{Psiscr}
\Psi^{\mathrm{scr}}_k \equiv \frac{ t_{de} \Gamma_k^+ q_{ek} - t_{do} \Gamma_k^- q_{ok} }{ \sqrt{t_{d-}^2 + t_{d+}^2} \sqrt{ |\Gamma_k^+|^2 \left( 1 + \gamma \cos \vphi \right) + |\Gamma_k^-|^2 \left( 1 - \gamma \cos \vphi \right) } }
\ee
where we have defined the asymmetry parameter
\be
\label{gamma}
\gamma \equiv \frac{2 t_{d-} t_{d+}}{t_{d-}^2 + t_{d+}^2}.
\ee

In this way the dot coupling Hamiltonian can be written as
\be
\label{Htdscreen}
H_{td} = \int_0^{\frac{\pi}{a}} dk \, \tilde{V}_{d k} \left( {\Psi^{\mathrm{scr}}_k}^\dag d + \hc \right)
\ee
where
\begin{widetext}
\bea
\tilde{V}_{dk} & \equiv &  - \sqrt{\frac{a}{\pi}} \sqrt{t_{d-}^2 + t_{d+}^2} \sqrt{ |\Gamma_k^+|^2 \left( 1 + \gamma \cos \vphi \right) + |\Gamma_k^-|^2 \left( 1 - \gamma \cos \vphi \right) } \\
\label{tildeVdk}
& = & - |\sin ka| \sqrt{\frac{2a}{\pi} \left( t_{d-}^2 + t_{d+}^2 \right) } \sqrt{ \frac{ 1 + {\tau^\p}^2 - 2 \gamma \tau^\p \cos \vphi \cos ka }{ (1 + {\tau^\p}^2)^2 - 4 {\tau^\p}^2 \cos^2 ka } }.
\eea
\end{widetext}
This form of the hybridization was first found in ref.~\onlinecite{yoshii2008}.

We are interested primarily in the Kondo effect which involves only the screening channel since it is the only one that couples to the quantum dot.  Hence, one can perform the Schrieffer-Wolff transformation~\cite{schrieffer1966, hewson1993} so that the screening channel Hamiltonian assumes the form
\bea
\nonumber
H & = & \int_0^{\frac{\pi}{a}} dk \, \eps_k {\Psi_k^\mathrm{scr}}^\dag \Psi_k^\mathrm{scr} \\
\label{fullHKondo}
& & + \int_0^{\frac{\pi}{a}} dk \, dk^\p \left( J_{k k^\p} \vec{s}_{k k^\p} \cdot \vec{S}_d + K_{k k^\p}  {\Psi_k^\mathrm{scr}}^\dag \Psi_{k^\p}^\mathrm{scr} \right)
\eea
where we have defined the screening channel spin density
\be
\vec{s}_{k k^\p} \equiv {\Psi_k^\mathrm{scr}}^\dag \frac{\vec{\sigma}}{2} \Psi_{k^\p}^\mathrm{scr}
\ee
and effective dot spin
\be
\vec{S}_d \equiv d^\dag \frac{\vec{\sigma}}{2} d
\ee
with $\vec{\sigma}$ being the three Pauli matrices (recall that $\Psi_k^\mathrm{scr}$ and $d$ are spinors).  The coupling parameters are given by
\bea
\label{fullSWJ}
J_{k k^\p} & = & \tilde{V}_{dk} \tilde{V}_{d k^\p} \left( \frac{1}{\eps_{k} - \eps_F - \eps_d} + \frac{1}{U + \eps_d - \eps_{k^\p} + \eps_F}  \right) \\
\label{fullSWK}
K_{k k^\p} & = & \frac{\tilde{V}_{d k} \tilde{V}_{d k^\p}}{2} \left( \frac{1}{\eps_k - \eps_F - \eps_d} - \frac{1}{U + \eps_d - \eps_{k^\p} + \eps_F} \right) \quad \quad 
\eea

At this point, one can obtain a low-energy effective theory by integrating out high-energy modes in the usual way.  The potential scattering term $K_{k k^\p}$ is marginal and does not renormalize.  We will discuss this term in more detail in \S~\ref{sec:kr} and neglect it for now.  The exchange interaction is relevant and diverges, giving rise to the usual Kondo screening of $\vec{S}_d$ by the screening channel Fermions for temperatures $T$ below the Kondo temperature $T_K$.  There are, however, physical consequences due to the flux $\vphi$ and inter-lead coupling $t^\p$ that will be determined in \S~\ref{sec:physprop}.  

To summarize the analysis thus far, through a series of basis rotations we have cast the inter-lead Hamiltonian into a potential scattering form.  By transforming to the scattering basis, we have eliminated this potential scattering term and identified the operator that couples directly to the quantum dot.  It is this combination that will participate in the Kondo screening of the dot.  Nevertheless, there are additional potential scattering terms that can arise in the screening channel and it is this subject that we next discuss.

\section{Additional Potential Scattering}
\label{sec:potscat}

Our goal is to derive an effective theory of our system that is valid at low temperatures, keeping the leading order contributions in the effective strength of the Kondo coupling 
\be
\label{Jdef}
J \equiv \frac{1}{a} J_{k_F k_F}|_{t^\p = 0} = \frac{2 \left(t_{d-}^2 + t_{d+}^2\right) \sin^2 k_F a}{\pi} \left( \frac{U}{-\eps_d \left(U + \eps_d \right)} \right)
\ee
which has dimensions of energy and which we take to be a small parameter.

The effective theory can be derived, to a first approximation, by linearizing the dispersion $\eps_k$ in a region $-Q < k - k_F < Q$  and approximating the coupling constants $J_{k k^\p}$ and $K_{k k^\p}$ by their values at the Fermi energy $J_{k_F k_F}$ and $K_{k_F k_F}$.  However, it will be shown that when the dot level is tuned to the value $\eps_d = -U/2$, $K_{k_F k_F}$ vanishes to second order in $V_{dk_F}$.  In this case, a more careful derivation of the low-energy Hamiltonian reveals that there is still an additional potential scattering generated by the renormalization of $J_{k k^\p}$.  This is higher order in $J$ than the leading order contribution to $K_{k_F k_F}$ written in eq.~(\ref{fullSWK}) but contributes to the leading order term in the additional potential scattering when eq.~(\ref{fullSWK}) vanishes at $\eps_d = -U/2$.  We address each of these cases separately below.

\subsection{Asymmetric dot $\eps_d \ne -U/2$}
\label{sec:kr}
Restricting excitations to a small region about the Fermi energy as described above, the potential scattering term generated by the Schrieffer-Wolff transformation assumes the form
\be
H_R = K_R \int_{-Q}^Q dk \, dk^\p \, {\Psi_{k}^{\mathrm{scr}}}^\dag \Psi_k^{\mathrm{scr}}
\ee
where
\begin{widetext}
\be
\label{KR}
K_R \equiv K_{k_F k_F} = \frac{a}{\pi} \sin^2 k_F a \left( t_{d-}^2 + t_{d+}^2 \right) \left( - \frac{U + 2\eps_d}{\eps_d(U + \eps_d)} \right) \frac{1 + {\tau^\p}^2 - 2 \gamma \tau^\p \cos \vphi \cos k_F a}{ (1 + {\tau^\p}^2)^2 - 4 {\tau^\p}^2 \cos^2 k_F a}.
\ee
\end{widetext}
In order to observe the Kondo effect, we require that $\eps_d \approx -U/2$ so as to favor the formation of a local moment rather than a doubly occupied or unoccupied dot level.  In this case, we see that $K_R$ is of order $J$.  However, for the precise value of $\eps_d = -U/2$, $K_R$ vanishes and there is no potential scattering generated directly by the Schrieffer-Wolff transformation at low energies to linear order in $J$.

The presence of this potential scattering term will give rise to a phase shift $\delta_R$ at the Fermi surface in the screening channel.  As shown in Appendix~\ref{sec:phaseshift}, this is given by 
\be
\label{dRKR}
\tan \delta_R = - \pi \nu K_R
\ee
for small $K_R$ and where $\nu$ is the density of states at the Fermi energy.  We will show in \S~\ref{sec:conductance} how this additional potential scattering contributes to the $T=0$ conductance of the AB ring.

\subsection{Symmetric dot $\eps_d = -U/2$}
\label{sec:vr}
As discussed above, integrating out the high-energy modes to obtain a low-energy Hamiltonian leaves the marginal interaction $K_{k k^\p}$ unchanged and so one obtains the term discussed in the above section.  However, one can ask the question as to whether or not an additional potential scattering term is generated by the Kondo interaction $J_{k k^\p}$.  Normally this is not the case for one often considers a Kondo interaction that is particle-hole symmetric.  It can be shown that this is not true for $J_{k k^\p}$ of eq.~(\ref{fullSWJ}).  This is a consequence of a non-zero $t^\p$ which necessarily breaks particle-hole symmetry.  Although we have transformed away the explicit $t^\p$ interaction, the particle-hole symmetry breaking is manifest in this more complicated Kondo interaction.  As a result, there is no symmetry forbidding this Kondo interaction from generating an \emph{additional} potential scattering term and it is to the calculation of this that we now turn our attention.  

Consider a renormalization group scaling by integrating out all of the wave vectors down to the Fermi energy.  Although it is difficult to perform such a transformation exactly, one can make progress through a perturbative expansion in $J$.  The leading order contribution is of order $J^2$ which will be much smaller than $K_R$, eq.~(\ref{KR}), which is of order $J$.  However, $K_R$ vanishes when $\eps_d = -U/2$ so that the $J^2$ term calculated below will contribute to the leading order term in the potential scattering.  Hence, in this section, we assume that $\eps_d = -U/2$.

Evaluating the Feynman diagrams to second order in the Kondo interaction $J_{k k^\p}$ in eq.~(\ref{fullHKondo}), one finds a potential scattering term generated of the form
\be
\label{HRscreening}
H_R = V_R \int_{-Q}^Q dk \, dk^\p \, {\Psi^{\mathrm{scr}}_k}^\dag \Psi^{\mathrm{scr}}_{k^\p} 
\ee
where the region of integration is restricted to small momentum about the Fermi momentum $k_F$ and $V_R$ is given by
\be
V_R = \frac{3}{16} \int_0^{\frac{\pi}{a}} dk \frac{J_{k_F k} J_{k k_F}}{\eps_F - \eps_k + i \eta \mathrm{sgn}(\eps_F - \eps_k)}.
\ee
The factor of $3/16$ comes from the trace over spin degrees of freedom and the denominator is simply the time-ordered propagator of the intermediate $\Psi_k^\mathrm{scr}$ Fermion.  

\begin{widetext}
Substituting in the definition of $J_{k k^\p}$ of eq.~(\ref{fullSWJ}) together with the definition of $\tilde{V}_{dk}$ from eq.~(\ref{tildeVdk}) and $J$ from eq.~(\ref{Jdef}), $V_R$ can be written as
\be
V_R = \frac{3a}{128 \sin^4 k_F a} \frac{J^2}{t} \left( |\Gamma_{k_F}^+|^2 (1+\gamma \cos \vphi) + |\Gamma_{k_F}^-|^2 (1-\gamma \cos \vphi) \right) \left( I_R^+ (1+\gamma \cos \vphi) + I_R^-(1-\gamma \cos \vphi) \right).
\ee
The factors of $I_R^\pm$ are dimensionless integrals given by
\be
I_R^\pm \equiv \int_0^\pi dy \frac{\sin^2 y}{1 \pm 2 \tau^\p \cos y + {\tau^\p}^2} \frac{1}{\cos y - \cos k_F a + i \eta \mathrm{sgn}(\cos y - \cos k_F a)} \frac{u^2 - (\cos y - \cos k_F)^2}{u^2 - 4 (\cos y - \cos k_F)^2} 
\ee
with
\be
u \equiv \frac{U}{2t}.
\ee

To evaluate these integrals, we break them up into two regions
\be
I_R^\pm = \left( \int_0^{k_F a} dy \frac{1}{\cos y - \cos k_F a + i \eta } + \int_{k_F a}^\pi dy   \frac{1}{\cos y - \cos k_F a - i \eta } \right)  \frac{\sin^2 y}{1 \pm 2 \tau^\p \cos y + {\tau^\p}^2}  \frac{u^2 - (\cos y - \cos k_F)^2}{u^2 - 4 (\cos y - \cos k_F)^2}.
\ee 
The imaginary parts from each integral cancel each other.  Upon evaluation of the principle part of each integral, one obtains
\be
I_R^\pm = \pm \frac{\pi}{8 \tau^\p} \left[ \frac{ 1 - {\tau^\p}^2 }{ 1 \pm 2 \tau^\p \cos k_F a + {\tau^\p}^2 } \left( 1 - \frac{3u^2 {\tau^\p}^2}{(1 \pm 2 \tau^\p \cos k_F a + {\tau^\p}^2)^2 - u^2 {\tau^\p}^2}  \right) - 1\right].
\ee 
Substituting this back into the above expression gives us our final result for $V_R$:
\bea
\nonumber
\nu V_R & = & - \frac{3 \pi^2 (\nu J)^2}{64 \tau^\p \sin k_F a} \frac{1 + {\tau^\p}^2 - 2 \gamma \tau^\p \cos k_F a \cos \vphi}{(1 + {\tau^\p}^2)^2 - 4 {\tau^\p}^2 \cos^2 k_F a} \\
\nonumber
& & \times \left\{ \gamma \cos \vphi + \frac{ 2 \tau^\p (1-{\tau^\p}^2) \cos k_F a - (1 - {\tau^\p}^4) \gamma \cos \vphi}{ (1 + {\tau^\p}^2)^2 - 4 {\tau^\p}^2 \cos^2 k_F a} \right. \\
\label{VR}
& & \qquad \left. + \frac{ 3 u^2 \gamma \cos \vphi \, {\tau^\p}^2 (1 - {\tau^\p}^2) \left[ (1 + {\tau^\p}^2)^2 + 4 {\tau^\p}^2 \cos^2 k_F a - u^2 {\tau^\p}^2 \right] - 12 u^2 {\tau^\p}^3 (1 - {\tau^\p}^4) \cos k_Fa}{ \left[ (1 + {\tau^\p}^2)^2 + 4 {\tau^\p}^2 \cos^2 k_F a - u^2 {\tau^\p}^2 \right]^2 - 16 {\tau^\p}^2 (1 + {\tau^\p}^2)^2 \cos^2 k_F a } \right\}
\eea
\end{widetext}
where $\nu$ is the density of states at the Fermi energy.  Just as with the potential scattering term $K_R$, this $V_R$ term will give rise to a phase shift in the screening channel given by
\be
\label{dRVR}
\tan \delta_R = - \pi \nu V_R
\ee
as shown in Appendix~\ref{sec:phaseshift}.

It should be noted that, although the potential scattering generated by the Schrieffer-Wolff transformation vanishes to order $J$, there may be a non-zero term at order $J^2$ in addition to that given by $V_R$ calculated above.  Such a calculation of the higher-order Schrieffer-Wolff terms is beyond the scope of this paper and so we leave it as a future project.  

In conclusion, the transformation analysis of section~\ref{sec:model} provides a simple, generic way to account for the presence of inter-lead coupling which takes the form of a potential scattering interaction.  Such a transformation effectively removes the potential scattering explicitly from the Hamiltonian in favor of a more complicated, particle-hole asymetric Kondo interaction when the dot is tuned to the Kondo regime.  We have further shown that additional potential scattering terms are generated in the screening channel.  The leading order contribution to this additional potential scattering is given by $K_R$, eq.~(\ref{KR}), in the case that $\eps_d \ne -U/2$ and by $V_R$, eq.~(\ref{VR}), when $\eps_d = -U/2$.   In the next section, we analyse the physical consequences of this low-energy model.

\section{Physical Properties}
\label{sec:physprop}

\subsection{Kondo Temperature}
\label{sec:TK}
One of the primary insights of the scattering transformation analysis is in revealing how the Aharanov-Bohm ring influences the coupling between the quantum dot and the screening channel of electrons.  That is, it allows us to obtain an expression for the dot-lead coupling in the Hamiltonian of eq.~(\ref{Htdscreen}), given by $\tilde{V}_{d k_F}$ (in the long wavelength limit), showing the dependence of the coupling on $t^\p$, $\vphi$, and $k_F$.   We then determine the $t^\p$, $\vphi$, and $k_F$ dependence of the effective Kondo coupling via the Schrieffer-Wolff transformation, eq.~(\ref{fullSWJ}).  This, in turn, gives rise to a $t^\p$, $\vphi$, and $k_F$ dependent Kondo temperature, the precise expression of which is easy to derive.  

\begin{figure}
\begin{center}
\includegraphics[width=0.45\textwidth, clip=true]{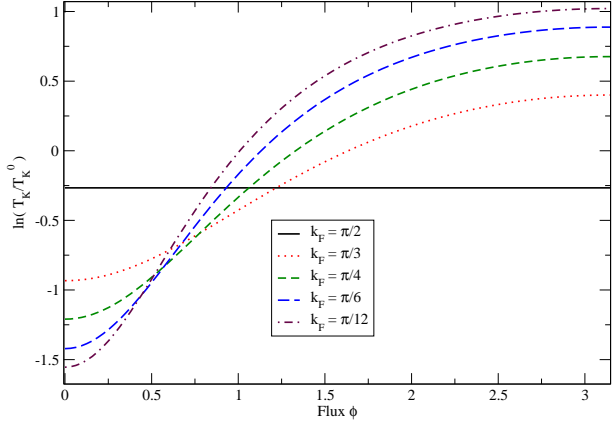}
\end{center}
\caption{\label{fig:TKvsphi} (Color online) The flux dependence of the Kondo temperature for a value of $\tau^\p = 0.4$ and $\gamma = 1$.  Here we see an increase in the flux dependence as the electron density is lowered.} 
\end{figure}

\begin{figure}
\begin{center}
\includegraphics[width=0.45\textwidth, clip=true]{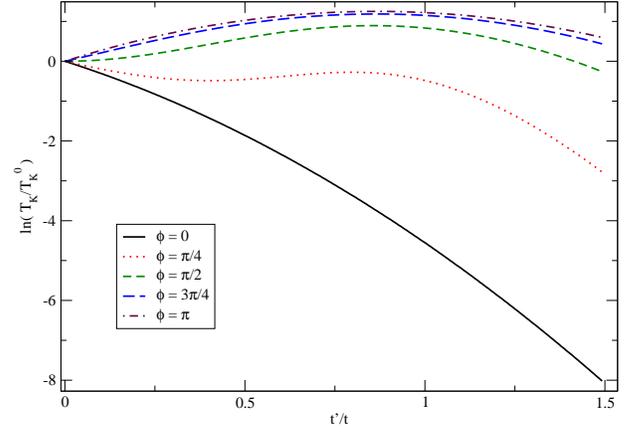}
\end{center}
\caption{\label{fig:TKvstp} (Color online) The $\tau^\p$ dependence of the Kondo temperature for a value of $k_F = \pi / 6a$.  This exhibits the variety of behaviour that can be seen for different values of the flux and that the Kondo temperature always vanishes as $\tau^\p \to \infty$.} 
\end{figure}

Using the low-energy effective Hamiltonian, we determine the effective Kondo coupling by evaluating eq.~(\ref{fullSWJ}) at the Fermi energy
\bea
J^{\mathrm{eff}} & \equiv & J_{k_F k_F} = \tilde{V}_{dk_F}^2 \frac{-U}{\eps_d \left( U + \eps_d \right)} \\
\label{effJ-AB} 
& = & J \frac{1 + {\tau^\p}^2 - 2 \gamma \tau^\p \cos \vphi \cos k_F a}{(1 + {\tau^\p}^2)^2 - 4 {\tau^\p}^2 \cos^2 k_F a}
\eea
where $J$ is defined in eq.~(\ref{Jdef}).   The leading order RG definition of the Kondo temperature~\cite{hewson1993} is
\be
T_K = D e^{-1/ (2 \nu J^{\mathrm{eff}})}
\ee
and dividing by the $t^\p=0$ Kondo temperature $T_K^0 = D e^{-1/(2 \nu J)}$, we get
\be
\label{lnTK}
\ln \frac{T_K}{T_K^0} = - \frac{\tau^\p}{2 \nu J} \frac{ 2 \gamma \cos \vphi \cos k_F a + \tau^\p \left( 1 - 4 \cos^2 k_F a \right) + {\tau^\p}^3}{1 - 2 \gamma \tau^\p \cos \vphi \cos k_F a + {\tau^\p}^2} .
\ee
Although the denominator is always positive, we see that the Kondo temperature can be raised or lowered by the presence of the Aharanov-Bohm ring depending on the values of $\tau^\p$, $\vphi$ and $k_F$.  This is shown in figures~\ref{fig:TKvsphi} and~\ref{fig:TKvstp} which show the flux and $t^\p$ dependence for various values of the other parameters.  

For the special case of half-filled leads, $k_F = \pi / (2a)$, the result is particularly simple
\be
\label{TKvstp}
\left. \ln \frac{T_K}{T_K^0} \right|_{k_F = \frac{\pi}{2a}} = - \frac{ {\tau^\p}^2}{2 \nu J} 
\ee
showing that the Kondo temperature is independent of flux in this case.  This limiting form of the Kondo temperature is verified by the NRG as discussed in \S~\ref{sec:nrg}.

\subsection{S-Matrix and Conductance}
\label{sec:conductance}
The strong-coupling fixed point of the Aharanov-Bohm model under consideration can be described by a two-channel Fermi liquid.  In this way, the fixed point model is fully described by a $2 \times 2$ S-matrix describing how the quasi-particle excitations of the two channels are scattered at the Fermi energy.  In this section, we derive this S-matrix and relate it to the conductance between the two leads.

The analysis of sections~\ref{sec:model} and~\ref{sec:potscat} provide the following simple picture of the strong-coupling fixed point.  The direct coupling between the two leads, $t^\p$, gives rise to a phase shift $\delta^\pm$ in the $q_{ek}$ and $q_{ok}$ channels respectively.  The form of these phase shifts is presented in eq.~(\ref{deltadef}) as computed in Appendix~\ref{sec:phaseshift}.  By transforming to the scattering basis and removing the $t^\p$ interaction from the Hamiltonian, we were able to identify the screening channel of eq.~(\ref{Psiscr}).  Defining the orthogonal complement, $\tilde{\Psi}_k^\mathrm{scr}$, to $\Psi_k^\mathrm{scr}$ and evaluating both at the Fermi energy  (relevant here since we are talking about $T=0$ properties), we can write the relation between the screening/non-screening basis and the even odd basis in terms of the above phase shifts as
\be 
\left(
\begin{array}{c}
\Psi^{\mathrm{scr}}_k \\
\tilde{\Psi}^{\mathrm{scr}}_k
\end{array}
\right) = U \left(
\begin{array}{c}
q_{ek} \\
q_{ok}
\end{array}
\right)
\ee
where
\begin{widetext}
\be
U = \mathcal{N} 
\left(
\begin{array}{cc}
-e^{-i \delta^+} \sin \delta^+ (t_{d-}e^{-i \povertwo} + t_{d+} e^{i \povertwo}) & -e^{-i \delta^-} \sin \delta^- (t_{d-}e^{-i \povertwo} - t_{d+} e^{i \povertwo}) \\
e^{i \delta^-} \sin \delta^- (t_{d-}e^{i \povertwo} - t_{d+} e^{-i \povertwo}) & -e^{i \delta^+} \sin \delta^+ (t_{d-}e^{i \povertwo} + t_{d+} e^{-i \povertwo})
\end{array}
\right)
\ee
with normalization
\be
\mathcal{N} \equiv \left[ (t_{d-}^2 + t_{d+}^2) ( \sin^2 \delta^+ (1 + \gamma \cos \vphi) + \sin^2 \delta^- (1 - \gamma \cos \vphi) ) \right]^{-\frac{1}{2}}.
\ee
\end{widetext}

In the screening channel, there will be a phase shift with two contributions.  The first is the usual $\pi / 2$ Kondo phase shift.  The second is the phase shift $\delta_R$ generated by the additional potential scattering, the leading order contribution to which will either be $K_R$ or $V_R$~\cite{footnote2} depending on the value of $\eps_d$.  Since the the additional potential scattering was obtained by integrating out the high-energy modes, the generated Hamiltonian term of eq.~(\ref{HRscreening}) must be considered as a low energy, long wavelength continuum model where the influence of the lattice is inconsequential.  The phase shift for such a model is derived in Appendix~\ref{sec:phaseshift} and shown to be either that of eq.~(\ref{dRKR}) or eq.~(\ref{dRVR}).

This is all of the information we require to write down the S-matrix in the even/odd basis:
\be
S = U^\dag \left(
\begin{array}{cc}
- e^{2 i \delta_R} & 0 \\
0 & 1
\end{array}
\right) U \left(
\begin{array}{cc}
e^{2 i \delta} & 0 \\
0 & e^{-2 i \delta}
\end{array}
\right).
\ee
The far right matrix describes the potential scattering phase shifts due to $t^\p$ in the $q_{ek}$ and $q_{ok}$ channels, $U$ rotates the basis to the screening channel and the matrix between $U$ and $U^\dag$ describes the phase shift $\delta_R$ due to $V_R$ or $K_R$ as well as the $\pi / 2$ Kondo phase shift giving rise to the factor of $-1 = e^{2 i \frac{\pi}{2}}$.  

Multiplying the matrices, we can write $S$ as
\be
\label{Smatrixeo}
S = \left(
\begin{array}{cc}
S_{ee} & S_{eo} \\
S_{oe} & S_{oo}
\end{array}
\right)
\ee
with
\begin{widetext}
\bea
S_{ee} & = & - \mathcal{M} e^{2i\delta^+} \left( e^{2i\delta_R}(1+\gamma \cos \vphi) \sin^2 \delta^+  - (1-\gamma \cos \vphi) \sin^2 \delta^-  \right) \\
S_{eo} &= & -2 \mathcal{M} e^{i(\delta^- + \delta^+ + \delta_R)} ( \beta - i \gamma \sin \vphi ) \sin \delta^- \sin \delta^+ \cos \delta_R  \\
S_{oe} & = & -2 \mathcal{M} e^{i(\delta^- + \delta^+ + \delta_R)} ( \beta + i \gamma \sin \vphi ) \sin \delta^- \sin \delta^+ \cos \delta_R  \\
S_{oo} & = & \mathcal{M} e^{2 i \delta^-} \left( (1+\gamma \cos \vphi) \sin^2 \delta^+  - e^{2i\delta_R} (1-\gamma \cos \vphi) \sin^2 \delta^-  \right)
\eea
\end{widetext}
where we have defined
\be
\beta \equiv \frac{t_{d-}^2 - t_{d+}^2}{t_{d-}^2 + t_{d+}^2}
\ee
and
\be
\mathcal{M} \equiv \frac{1}{(1+\gamma \cos \vphi) \sin^2 \delta^+  + (1-\gamma \cos \vphi) \sin^2 \delta^- }.
\ee

To relate this S-matrix to the conductance, we first construct general scattering wave functions between the even and odd channels.  Consider an incoming plane wave in the even channel that is then scattered into the even and odd outgoing channel according to the above S-matrix.  Such a wave function takes the form
\be
\psi_e = e^{-i k |x|} + S_{ee} e^{i k |x|} + S_{oe} \sgn(x) e^{i k |x|} 
\ee
where the first term is the incoming wave in the even channel, the second term the scattered even wave and the last term the scattered odd wave.  Similarly, considering an incoming wave in the odd channel gives the wave function
\be
\psi_o = \sgn(x) e^{-ik|x|} + S_{eo} e^{i k |x|} + S_{oo} \sgn(x) e^{i k |x|}.
\ee

Next, we wish to form a combination of $\psi_e$ and $\psi_o$ that corresponds to a right-moving wave incoming from the left.  That is, we wish to form a superposition of the above two wave functions that has \emph{no left-moving component for $x>0$}.  To this end, we form
\bea
\nonumber
\psi & \equiv & \frac{1}{2} \left( \psi_e - \psi_o \right) \\
& = & \left\{ 
\begin{array}{ll}
\frac{1}{2} \left( S_{ee} + S_{oe} - S_{eo} - S_{oo} \right) e^{ikx} & , x > 0 \\
e^{ikx} + \frac{1}{2} \left( S_{ee} + S_{oo} - S_{eo} - S_{oe} \right) e^{-ikx} & , x<0 \quad \, \, 
\end{array}
\right.
\eea
where, indeed, we find no $e^{-ikx}$ component in $\psi$ for $x>0$. 

Looking at the $x>0$ portion of $\psi$, we recognize the coefficient of the plane wave as the transmission probability amplitude for transmission from the left lead to the right lead
\be
T = \frac{1}{2} \left( S_{ee} + S_{oe} - S_{eo} - S_{oo} \right).
\ee
Using the Landauer-Buttiker formula, we obtain an expression for the conductance
\begin{widetext}
\bea
\nonumber
G & = & \frac{2 e^2}{h} |T|^2 \\
\nonumber
\label{LBcond} & = & \frac{2 e^2}{h} \biggl\{ (1+\gamma \cos \vphi)^2 \sin^4 \delta^+ \cos^2(\delta^+ - \delta^- + \delta_R)  + (1-\gamma \cos \vphi)^2 \sin^4 \delta^- \cos^2(\delta^+ - \delta^- - \delta_R)  \\
\nonumber
& & \qquad + \sin^2 \delta^- \sin^2 \delta^+ \left[ 4 \cos^2 \delta_R \sin^2 \vphi -  \left( 1 - \gamma^2 \cos^2 \vphi \right) \left( \cos \left[2 (\delta^+ - \delta^-) \right] + \cos 2 \delta_R \right)  \right] \\
\label{Gfull}
& & \qquad \biggr\} \bigg/ \left[  (1+\gamma \cos \vphi) \sin^2 \delta^+  + (1-\gamma \cos \vphi) \sin^2 \delta^- \right]^2
\eea
\end{widetext}
This is the most general expression for the conductance expressed in terms of the phase shifts $\delta^\pm$ generated by the inter-lead coupling $t^\p$, the additional potential scattering $K_R$ or $V_R$ via $\delta_R$, and in terms of the flux $\vphi$.  The latter includes the explicit $\vphi$ dependence written above as well as the dependence implicit in $\delta_R$ via the flux dependence of $K_R$ or $V_R$ written in eqs.~(\ref{KR}) or~(\ref{VR}).  Although the equation is rather complicated, we see that the conductance satisfies the necessary symmetry relation $G(\vphi) = G(-\vphi)$.  We now turn our attention to special limiting cases.

\begin{figure}
\begin{center}
\includegraphics[width=0.45\textwidth, clip=true]{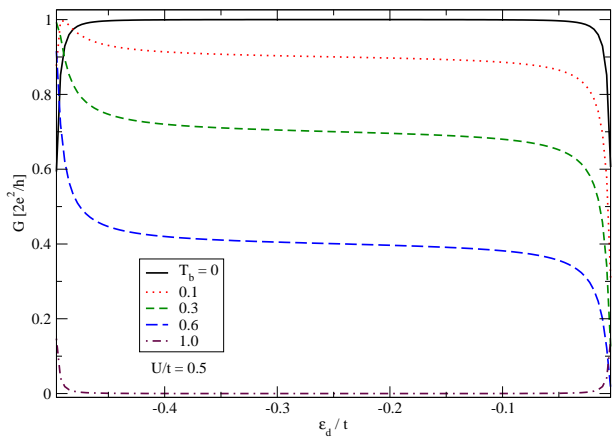}
\end{center}
\caption{\label{fig:Gvsepsd} (Color online) The conductance as a function of dot level $\eps_d$ for various values of the direct interlead transmission probability give by $T_b = \sin^2 2 \delta$.  Here, we assume the particle-hole symmetric value of half-filling, $k_F = \pi / 2a$ and $\gamma = 1$.  This is the behaviour seen numerically in ref.~\onlinecite{hofstetter2001}.} 
\end{figure}

For the case of $k_F = \pi / (2a)$ and $t_{d-} = t_{d+}$ considered in most previous studies, $\delta^+ = - \delta^- \equiv \delta$ with $\tan \delta = \tau^\p$ and the conductance simplifies to 
\begin{widetext}
\be
\label{Ghalf-filling}
\left. G \right|_{k_F = \frac{\pi}{2a}} =  \frac{2 e^2}{h} \left[ \cos^2(2 \delta - \delta_R) \cos^4 \povertwo + \cos^2(2 \delta + \delta_R) \sin^4 \povertwo  + \cos^2 \delta_R \sin^2 \vphi - \frac{1}{4} \left( \cos 4 \delta + \cos 2 \delta_R \right) \sin^2 \vphi \right].
\ee
\end{widetext}
It is interesting to compare this with the numerical results of ref.~\onlinecite{hofstetter2001}.  For the case of $\eps_d \ne -U/2$, when $K_R$ is the leading order contribution to $\delta_R$, we are able to qualitatively reproduce the Fano-Kondo behaviour seen in ref.~\onlinecite{hofstetter2001} in the region $\eps_d \approx -U/2$ for which our analysis is valid.  An example of this is given in figure~\ref{fig:Gvsepsd}.  

For the symmetric value $\eps_d = -U/2$ when $K_R$ vanishes, we can view the $\delta_R$ generated by $V_R$ as a small correction to the results of ref.~\onlinecite{hofstetter2001}.  Indeed, in the limit of $\delta_R \to 0$ and $k_F = \pi / (2a)$, our result reduces to
\be
\left. G \right|_{\delta_R = 0, k_F = \frac{\pi}{2a}} = \frac{2e^2}{h} \left( 1 - T_b \cos^2 \vphi \right)
\ee
where $T_b = \sin^2 2 \delta$ is the transmission probability through the lower arm of the Aharanov-Bohm ring in the absence of the upper arm.  This is precisely the form reported in [\onlinecite{hofstetter2001}] for the case of a singly-occupied quantum dot.  

In this way, eq.~(\ref{Ghalf-filling}) can be viewed as an analytic description of the results of ref.~\onlinecite{hofstetter2001}, the latter of which required numerical input from the NRG.  Such an analytic description is only valid for values of $\eps_d$ close to $-U/2$ so as to strongly favor a local moment on the quantum dot whereas the results of ref.~\onlinecite{hofstetter2001} are valid for all $\eps_d$.  On the other hand, our complete expression for the conductance, eq.~(\ref{Gfull}), extends previous results to cases where the Fermi energy is not situated in a particle-hole symmetric manner relative to the band edges (\textit{e.g.}\ $k_F \ne \pi / (2a)$) as well as taking into account the additional potential scattering $V_R$ discussed in \S~\ref{sec:potscat}.

\begin{figure}
\begin{center}
\includegraphics[width=0.45\textwidth, clip=true]{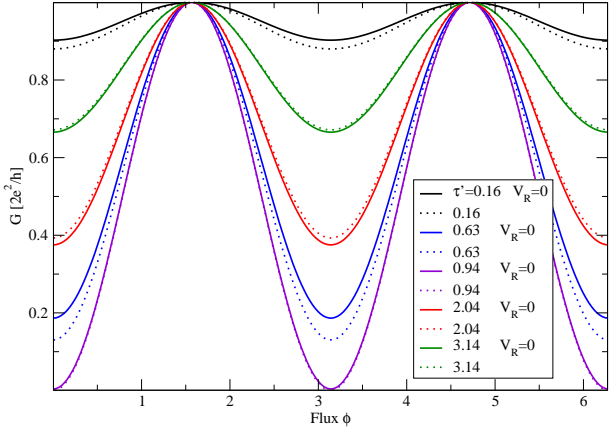}
\end{center}
\caption{\label{fig:Gflux} (Color online) The conductance plotted as a function of magnetic flux for $k_F = \pi / 2a$, $\eps_d = -U/2$ and $\gamma = 1$.  Each of the different coloured lines indicates a different value of the inter-lead coupling $\tau^\p$.  The solid lines are the prediction with $V_R = 0$ (equivalently $\delta_R = 0$) with the dotted lines showing the finite $V_R$ correction.  This data assumes a value $\nu J = 0.287$ for the bare Kondo coupling. } 
\end{figure}
\begin{figure}
\begin{center}
\includegraphics[width=0.45\textwidth, clip=true]{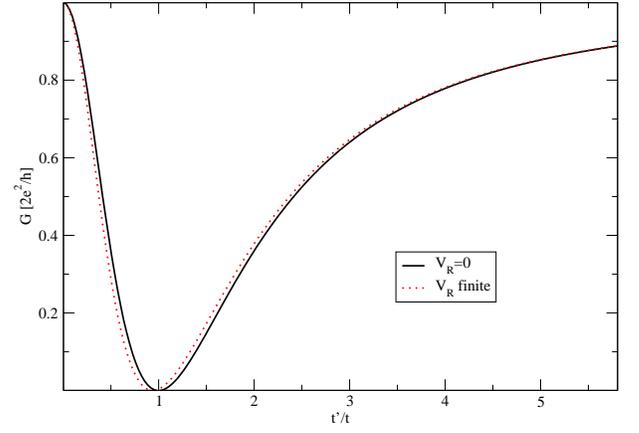}
\end{center}
\caption{\label{fig:GVp} (Color online) The conductance plotted as a function of inter-lead coupling $t_p$ when $k_F = \pi / 2a$ and $\eps_d = -U/2$.  Here, $J = 0.287$ and $\vphi = 0$.} 
\end{figure}
To further examine the correction due to $V_R$, we look at the flux dependence of the conductance in figure~\ref{fig:Gflux} for the case that $\eps_d = -U/2$ and hence $V_R$ contributes to the leading order behaviour of $\delta_R$.  There, each of the different coloured lines indicates a different value of the direct inter-lead coupling $t^\p$ as encoded by $\delta$.  It is seen that the conductance contrast (the difference between the minimum and maximum conductance) reaches a maximum for an intermediate value of the inter-lead coupling $\tau^\p = 1$ ($t^\p = t$).  

Furthermore, it is shown that for $\tau^\p < 1$, the effect of the additional potential scattering $V_R$ is to \emph{decrease} the conductance whereas for values $\tau^\p > 1$, the additional potential scattering serves to \emph{increase} the conductance.  This fact is made more evident in figure~\ref{fig:GVp} where the conductance is plotted versus $\tau^\p$ for $\vphi = 0$.  There, one can clearly see the crossover from reduced to enhanced conductance around $\tau^\p = 1$.  

Given that $V_R$ offers only a small correction, we look at the $\delta_R = 0$ limit of the conductance for general $k_F$ which takes the form
\begin{widetext}
\bea
\nonumber
\left. G \right|_{\delta_R = 0}  & = & \frac{2 e^2}{h} \biggl\{ \cos^2(\delta^+ - \delta^-) \left( \sin^4 \delta^+  (1+\gamma \cos \vphi)^2 + \sin^4 \delta^-  (1-\gamma \cos \vphi)^2 \right) \\
\nonumber
& & \qquad + \sin^2 \delta^- \sin^2 \delta^+ \left[ 4 \sin^2 \vphi - 2 \cos^2(\delta^+ - \delta^-) \left( 1 - \gamma^2 \cos^2 \vphi \right) \right] \\
& & \qquad \biggr\} \bigg/ \left[  \sin^2 \delta^+ (1+\gamma \cos \vphi) + \sin^2 \delta^- (1-\gamma \cos \vphi)\right]^2
\eea
\end{widetext}
Even without including the small correction due to $V_R$, this is a generalization of the conductance reported in ref.~\onlinecite{hofstetter2001} which, like most similar studies, only considered the case where the leads exhibit particle-hole symmetry ($k_F = \pi / (2a)$ for our tight-binding leads).  For quantum dots constructed on semi-conductor heterostructures where the two-dimensional electron gas has very low density, the Fermi energy will be very close to the bottom of the energy band and so exhibit strong particle-hole asymmetry.  Hence, the generalized forms for the conductance reported above seem to be more applicable to such devices than those reported in previous studies.

The description of the conductance that emerges from this analysis is quite interesting.  In the limit that $\tau^\p \to 0$, we recover the well-studied model of a single quantum dot embedded between two leads where one obtains unitary conductance at zero temperature.  As one increases $\tau^\p$, interference effects play a stronger role until one obtains maximal interference at $\tau^\p = 1$ ($T_b = 1$) where one is able to obtain total destructive interference in the form of zero conductance for certain values of the parameters (\textit{e.g.}\  $k_F = \pi / (2a)$ and $\vphi = 0$).  As one further increases $\tau^\p$, the transmission $T_b$ through the lower arm decreases and interference effects are diminished.  

\section{Support from the Numerical Renormalization Group}
\label{sec:nrg}

\subsection{Phase Diagram \& Kondo Temperature}

\subsubsection{Fixed points of the single-channel Anderson impurity model}
We begin by reviewing the various fixed points present in the single-channel Anderson model~\cite{krishnamurthy1980a} before describing the influence of the Aharanov-Bohm ring.  When investigating low-energy, long-wavelength properties, it is customary to define a model in terms of continuous fields with a linearized dispersion relation characterized by a Fermi velocity $v_F$.  In this way, we can write the single-channel Anderson model in terms of right-moving 1D electron annihilation operators $\psi(x)$ as
\bea
\nonumber
H & = & v_F \int_{-\infty}^\infty dx \, \psi^\dag(x) \left( - i \partial_x \right) \psi(x) + V_d \left( \psi^\dag(0) d + \hc \right) \\
& & + \frac{U}{2} \left( d^\dag d - 1 \right)^2 - \frac{U}{2}
\eea
where we have set the dot level to $\eps_d = - U /2$ (assumed throughout this section).  

\begin{table}
\begin{center}
\begin{tabular}{l|ccl}
\hline
\textbf{Fixed Point} & $V_d$ & $U$ & Stability\\
\hline
Free Orbital (FO) & 0 & 0 & Unstable\\
Local Moment (LM) & 0 & $\infty$ & Unstable\\
Strong Coupling (SC) & $\infty$ & $\infty$ & Stable\\
\hline
\end{tabular}
\end{center}
\caption{\label{table:AMFP} Summary of fixed points for the single-channel Anderson impurity model.}
\end{table}
This model has three fixed points summarized in Table~\ref{table:AMFP}.  The \emph{free orbital (FO) fixed point} occurs when $V_d = U = 0$.  This describes free $\psi$ Fermions with a decoupled free dot level $d$.  The spectrum of such a model is that of free Fermions plus the \emph{four} degenerate, zero-energy states of the dot.  

The FO fixed point is unstable and flows towards the \emph{local moment (LM) fixed point} as the energy scale is lowered.  The LM fixed point is characterized by a diverging $U \to \infty$ and $V_d = 0$.  This LM fixed point is the same as the FO except that two of the four dot levels are energetically forbidden, namely, those for which $d^\dag d = 0$ and $d^\dag d = 2$.  In other words, the quantum dot can only be singly occupied with either a spin up or spin down electron.  Hence, the spectrum will be that of free Fermions plus \emph{two} degenerate, zero-energy states of the dot.

The LM fixed point is also unstable and eventually flows to the \emph{strong coupling (SC) fixed point} described by a diverging $|V_d|^2 / U \to \infty$.  The nature of this fixed point can most easily be understood by first considering a Hamiltonian close to the LM fixed point with a small $|V_d| \ll U$.  In this case, one can perform a Schrieffer-Wolff transformation~\cite{schrieffer1966} perturbatively in $V_d$ to obtain a dot interaction 
\be
H_{td} + H_{\mathrm{dot}} \approx J \psi^\dag(0) \frac{\vec{\sigma}}{2} \psi(0) \cdot \vec{S_d}
\ee
where $\bs{\sigma}$ is a vector of the three Pauli matrices, $\bs{S_d} \equiv d^\dag (\bs{\sigma}/2) d$ is the effective spin of the singly-occupied dot level, and the coupling strength $J$ is proportional to $|V_d|^2 / U$.  This is the Kondo interaction between the localized spin of the quantum dot and the electrons in the leads.  The SC fixed point of the Anderson model is essentially the same as the strong-coupling fixed point of the Kondo model wherein $J \to \infty$ and the dot spin is screened by forming a singlet with the lead electrons.  

\subsubsection{Fixed points of the Aharanov-Bohm quantum dot model}
\label{sec:FPABmodel}
The low-energy transformations of section~\ref{sec:model} reveals that the renormalization group flow for the Aharanov-Bohm ring model under consideration will be very similar to that of the single-channel Anderson model just described.  Indeed, we have learned that a single, independent combination of the lead electrons, $\Psi^{\mathrm{scr}}_k$, couples directly to the dot just as in the single-channel Anderson model.  The precise nature of this screening channel will depend on both flux $\vphi$ and the inter-lead coupling $t^\p$ but the point is that there is a single channel available to screen the spin of the quantum dot.  For simplicity, we consider only the symmetric case where $t_{d-} = t_{d+} = t_d$ and $\eps_d = -U/2$.

The primary difference with the Anderson model discussed in the previous section is the addition of some potential scattering phase shifts $\delta^\pm$ depending on $t^\p$ and the modification of the dot-lead coupling $V_{d k_F} \to \tilde{V}_{d k_F}$.

We find that the FO and LM fixed points, with $\tilde{V}_{d k_F} = 0$, will be the same as in the single-channel Anderson model \emph{with the addition} of the phase shifts $\delta^\pm$ arising from the direct tunneling between the leads which were incorporated into the definition of $q_{ek}$ and $q_{ok}$.  The SC fixed point of the Aharanov-Bohm ring model will be one in which the dot spin is fully screened by the $\Psi^{\mathrm{scr}}_k$ combination of lead electrons.  Just as in the Kondo model, this will give rise to a $\pi/2$ phase shift in the $\Psi^{\mathrm{scr}}_k$ channel in addition to the phase shifts $\delta^\pm$ arising from the direct tunneling $t^\p$.   

Furthermore, the FO and LM fixed points occur for $\tilde{V}_{d k_F} = 0$ and, since $\tilde{V}_{d k_F}$ encodes the $t^\p$ dependence of the model, we predict that the cross-over scale of these fixed points will be unaffected by the presence of the Aharanov-Bohm ring (\textit{i.e.}~in the region of these fixed points, the $t^\p$ and $\vphi$ dependence of $\tilde{V}_{d k_F}$ is inconsequential).   However, the cross-over energy scale to the SC fixed point, that is, the Kondo temperature $T_K$, \emph{will} be influenced by the direct tunneling $t^\p$ and flux $\vphi$ as discussed in \S~\ref{sec:TK}.  

We can check these predictions for the fixed points of the Aharanov-Bohm ring model using the Numerical Renormalization Group (NRG).   This numerical algorithm is exhaustively detailed in the pioneering papers~\cite{wilson1975, krishnamurthy1980a, krishnamurthy1980b} and in a recent review~\cite{bulla2008} so we give only an outline sketch here.  

We begin with a long wavelength version of the Hamiltonian described in eqs.~(\ref{H0eo})--(\ref{Htdeo}) with a dispersion relation linearized about the Fermi energy $\eps_F = 0$, a cutoff in momentum at $k = \pm Q$ (here, $k$ is measured with respect to $k_F$, \textit{i.e.}\ $k-k_F \to k$), $t_{d-} = t_{d+} = t_d$, and $\eps_d = -U/2$.

The resulting Hamiltonian is
\begin{widetext}
\bea
\nonumber
H & = & v_F \int_{-Q}^Q dk \, k \left( e_k^\dag e_k + o_k^\dag o_k \right) - V_p \int_{-Q}^Q dk \, dk^\p \left( e_k^\dag e_{k^\p} - o_k^\dag o_{k^\p} \right) \\
\label{HcontNRG}
& & + V_d \int_{-Q}^Q dk \left[ \left( \cos \povertwo e_k^\dag + \sin \povertwo o_k^\dag \right) d + \hc \right] + \frac{U}{2} \left( d^\dag d - 1 \right)^2
\eea  
\end{widetext}
where we have simplified our notation by defining the potential scattering term $V_p \equiv - v_{k_F k_F}$, $V_d \equiv -2 t_d \sqrt{a/\pi} \sin k_Fa$, and redefining the phase of $o_k$ so as to make all coefficients real.  Note that this version of the Hamiltonian does \emph{not} involve a transformation to scattering states.  In this way, agreement between the NRG and results inferred from the transformations of section~\ref{sec:model} will serve as support for the scattering transformation analysis.

However, it should be observed that such a linear dispersion necessarily exhibits particle-hole symmetry whereas the tight-binding model discussed in section~\ref{sec:model} generally breaks particle-hole symmetry except for the special case of $k_F = \pi / (2a)$ that occurs when there is one electron per site.  For this reason, the NRG as formulated here strictly serves only to support our scattering transformation analysis for the particle-hole symmetric case of $k_F = \pi / (2a)$.  Nevertheless, we trust that our analytic results hold true for arbitrary $k_F$.

Setting up the NRG involves a series of transformations and approximations that map the model for the lead electrons onto two semi-infinite tight-binding chains, often termed Wilson chains, with hopping amplitudes that exponentially decrease with distance from the quantum dot
\bea
\label{HNRGinf}
\nonumber \frac{H}{D} & \approx & \frac{1}{2} \left( 1 + \Lambda^{-1} \right)^{\frac{1}{2}} \sum_{b = e, o} \sum_{n=0}^\infty \Lambda^{-\frac{n}{2}} \xi_n \left( f_{nb}^\dag f_{(n+1)b} + \hc \right) \\
\nonumber & &+ \frac{U}{2D} \left(d^\dag d - 1\right)^2 - 2 \nu V_p \left( f_{0e}^\dag f_{0e} - f_{0o}^\dag f_{0o} \right) \\
& & + \sqrt{\frac{2 \Gamma}{\pi D}} \left[ \left( \cos \frac{\vphi}{2} f_{0e}^\dag + \sin \frac{\vphi}{2} f_{0o}^\dag \right) d + \hc \right].
\eea
In general, each Fermionic $f_{n e}$ and $f_{n o}$ is a complicated linear combination of $e_k$ and $o_k$ respectively.  The details of this relationship are not of great importance for the present discussion except to note that the Fermions created on the $n=0$ site by $f_{0e}^\dag$ and $f_{0o}^\dag$ are proportional to the $e$ and $o$ electrons at the origin: $f_{0e} \propto e(x=0)$ and $f_{0o} \propto o(x=0)$.  Of the other parameters defined in this Wilson-chain Hamiltonian, $2D$ is the bandwidth and $\Lambda > 1$ is a dimensionless discretization parameter defined such that the continuum limit is recovered in the limit $\Lambda \to 1$.  The dimensionless parameter $\xi_n$ is given by
\be
\label{xin}
\xi_n = (1 - \Lambda^{-n-1}) (1 - \Lambda^{-2n-1})^{-1/2} (1 - \Lambda^{-2n-3})^{-1/2}
\ee
and tends to unity for $n \gg 1$.  We have also defined
\be
\label{Gammadef}
\Gamma \equiv  2 \pi \nu V_d^2 
\ee
with $\nu$ the density of states at the Fermi energy.

The renormalization group is realized by truncating the infinite chain to $N$ sites and rescaling the Hamiltonian such that the eigenvalues are of order unity
\bea
\nonumber H_N & \equiv \Lambda^{(N-1)/2} \biggl\{  & \sum_{b = e, o} \sum_{n=0}^{N-1} \Lambda^{-\frac{n}{2}} \xi_n \left( f_{nb}^\dag f_{(n+1)b} + \hc \right)  \\
\nonumber & & + \tilde{U} \left(d^\dag d - 1\right)^2 - \tilde{V}_p \left( f_{0e}^\dag f_{0e} - f_{0o}^\dag f_{0o} \right) \\
\nonumber
& &  + \tilde{\Gamma}^{\frac{1}{2}} \left[ \left( \cos \frac{\vphi}{2} f_{0e}^\dag + \sin \frac{\vphi}{2} f_{0o}^\dag \right) d \right. \\
\label{HN} 
& & \qquad \, \, \,  + \hc \Bigr]  \biggr\}
\eea
where the quantities with tildes are simply dimensionless versions of the original parameters of eq.~(\ref{HNRGinf}) with $\Lambda$ dependent rescaling.
The renormalization group transformation then takes the form of the recursion relation
\be
\label{NRGRGtrans}
H_{N+1} = \Lambda^{\frac{1}{2}} H_N + \xi_N \sum_{b=e, o} \left( f_{Nb}^\dag f_{(N+1)b} + \hc \right)
\ee
and is realized by iterative diagonalization, using the eigenvalues and eigenvectors of $H_N$ to define $H_{N+1}$ via eq.~(\ref{NRGRGtrans}).  In practice, the eigenvalues are shifted so that the lowest one is zero.  

The finite Hamiltonian $H_N$ can be related to the Hamiltonian of eq.~(\ref{HNRGinf}) by
\be
\frac{H}{D} = \lim_{N \to \infty} \frac{1}{2} \left( 1 + \Lambda^{-1} \right) \Lambda^{- (N-1)/2} H_N.
\ee
Since the dimensionless scale of $H_N$ is of order unity by definition, this indicates that the spectrum of $H_N$ describes the spectrum of the physical Hamiltonian at an energy scale given by
\be
\label{EN}
E_N \approx \frac{1}{2} \left(1 + \Lambda^{-1}\right) \Lambda^{- (N-1)/2} D.
\ee
In this way, we can associate $H_N$ with the effective Hamiltonian at the renormalization group energy scale $E_N$.   Fixed points can be identified as regions of $N$ over which the energy spectrum of the associated $H_N$ changes very little (for unstable fixed points) or not at all (for stable fixed points).  These fixed point NRG spectra can then be compared with that predicted by the scattering transformation analysis described above to test the validity of said analysis.

Our analysis of the fixed points follows that of~\cite{krishnamurthy1980a, krishnamurthy1980b}.  Let us first consider the FO fixed point which, in terms of the NRG formalism, is defined by $\tilde{\Gamma} = 0$ and $\tilde{U} = 0$, resulting in
\bea
\nonumber
H_{N, \mathrm{FO}} & = & \Lambda^{(N-1)/2} \biggl\{  \sum_{b = e, o} \sum_{n=0}^{N-1} \Lambda^{-\frac{n}{2}} \xi_n \left( f_{nb}^\dag f_{(n+1)b} + \hc \right) \\
\label{HNFO}
& & \qquad \qquad - \tilde{V_p} \left( f_{0e}^\dag f_{0e} - f_{0o}^\dag f_{0o} \right) \biggr\}.
\eea
This has the form of two decoupled Wilson chains, each with a potential scattering term at the origin.  Such chains were analyzed in~\cite{krishnamurthy1980b} where the $\tilde{V}_p$ dependence of the single-particle energies was described in detail.  

Extending their analysis to two decoupled channels as described in eq.~(\ref{HNFO}), one can diagonalize the non-interacting fixed point Hamiltonian and write it in terms of the single-particle and hole excitations
\begin{widetext}
\be
\label{HNFOdiag} H_{N, \mathrm{FO}} = \left\{ 
\begin{array}{ll}
\sum_{b=e, o} \sum_{n=1}^{(N+1)/2} \left( \eta^+_{nb}(\tilde{V}_p) g_{nb}^\dag g_{nb} + \eta^-_{nb}(\tilde{V}_p) h_{nb}^\dag h_{nb} \right) & \quad, N \, \textrm{odd} \\
\sum_{b=e, o} \left[ \sum_{n=1}^{N/2} \left( \hat{\eta}^+_{nb}(\tilde{V}_p) g_{nb}^\dag g_{nb} + \hat{\eta}^-_{nb}(\tilde{V}_p) h_{nb}^\dag h_{nb} \right) + \hat{\eta}^+_{0b} g_{0b}^\dag g_{0b} \right] & \quad, N \, \textrm{even}.
\end{array}
\right.
\ee 
\end{widetext}
Here, $g_{nb}$ destroys a quasiparticle while $h_{nb}$ destroys a quasihole.  The corresponding single particle/hole excitations are $N$-dependent in general but, for $N > 10$ (approximately), they are found to only depend on whether $N$ is even or odd, in which case one obtains $\tilde{V}_p$-dependent energy levels $\hat{\eta}^\pm(\tilde{V}_p)$ or $\eta^\pm(\tilde{V}_p)$ respectively.  

The precise numerical values of these energy levels depend on $\Lambda$ and $\tilde{V}_p$.  The $\tilde{V}_p$ dependence is described in~\cite{krishnamurthy1980b} where it was found that
\be
\eta^+_{nb}(\tilde{V}_p) = \eta^-_{nb}(-\tilde{V}_p)
\ee
and similarly for $\hat{\eta}$.  Furthermore, since the potential scattering in the $e$ channel is equal in magnitude but opposite in sign to that in the $o$ channel, the above relation can be written as
\be
\label{spEnergyRelation}
\eta^\pm_{ne}(\tilde{V}_p) = \eta^\mp_{no}(\tilde{V}_p)
\ee
and similarly for $\hat{\eta}$.  In this way, we recover a form of particle-hole symmetry even at finite $\tilde{V}_p$ where the energy spectrum of particles in the $e$ channel are equivalent to the spectrum of holes in the $o$ channel and \textit{vice versa}.

We can now combine these single-particle/hole excitations in multi-particle/hole combinations (being sure to respect the Pauli exclusion principle), together with the four degenerate zero-energy states of the dot level and so construct the FO fixed point spectrum.  The lowest such energy levels are given in Table~\ref{table:FOspec} along with the corresponding total charge $Q$ and total spin $S$ quantum numbers.  
\begin{table}
\begin{center}
\begin{tabular}{|cccc|cccc|}
\hline
Energy & Num.~Value & $Q$ & $2S$ & Energy & Num.~Value & $Q$ & $2S$ \\
\hline
0 & 0.0000 & -1 & 0 & $2\eta_{1e}^+ + \eta_{1o}^-$ & 0.4485 & 0 & 1 \\
& & 0 & 1 & & & 1 & 0 \\
& & 1 & 0& & & 1 & 2 \\
\cline{1-4}
$\eta_{1e}^+$ & 0.1495 & 0 & 1 & & & 2 & 1 \\
& & 1 & 0 & $\eta_{1e}^+ + 2\eta_{1o}^-$ & & 0 & 1 \\
& & 1 & 2 & & & -1 & 0 \\
& & 2 & 1 & & & -1 & 2 \\
$\eta_{1o}^-$ & & 0 & 1 & & & -2 & 1 \\
\cline{5-8}
& & -1 & 0 & $2\eta_{1e}^+ + 2\eta_{1o}^-$ & 0.5980 & -1 & 0\\
& & -1 & 2 & & & 0 & 1\\
& & -2 & 1 & & & 1 & 0 \\
\hline
$2 \eta_{1e}^+$ & 0.2990 & 1 & 0 & $\eta_{1o}^+$ & 1.3580 & 0 & 1\\
& & 2 & 1 & & & 1 & 0\\
& & 3 & 0 & & & 1 & 2\\
$2 \eta_{1o}^-$ & & -1 & 0 & & & 2 & 1\\
& & -2 & 1 & $\eta_{1e}^-$ & & 0 & 1\\
& & -3 & 0 & & & -1 & 0\\
$\eta_{1e}^+ + \eta_{1o}^-$ & & -1 & 0 & & & -1 & 2\\
& & -1 & 2 & & & -2 & 1\\
\cline{5-8}
& & 0 & 1 & & & & \\
& & 0 & 1 & & & & \\
& & 0 & 3 & & & & \\
& & 1 & 0 & & & & \\
& & 1 & 2 & & & & \\
\hline
\end{tabular}
\end{center}
\caption{\label{table:FOspec} The lowest energies and associated total charge $Q$ and total spin $S$ quantum numbers of the free orbital (FO) NRG fixed point of the Aharanov-Bohm ring model for odd $N$.  The numerical values for the single-particle excitation energies were obtained by diagonalizing the Hamiltonian of eq.~(\ref{HNFO}) using a value of $\tilde{V}_p = 3.0$ and $\Lambda = 2.5$.  All energies within a particular box are equal by eq.~(\ref{spEnergyRelation}).  }
\end{table}

The spectrum for the LM fixed point is closely related to that of the FO.  The corresponding NRG Wilson-chain Hamiltonian for the LM fixed point is
\bea
\nonumber H_{N, \mathrm{LM}} & = \Lambda^{(N-1)/2} \biggl\{ &  \sum_{b = e, o} \sum_{n=0}^{N-1} \Lambda^{-\frac{n}{2}} \xi_n \left( f_{nb}^\dag f_{(n+1)b} + \hc \right) \\
\nonumber
& & - \tilde{V_p} \left( f_{0e}^\dag f_{0e} - f_{0o}^\dag f_{0o} \right) \\
\label{HNLM} & & + \lim_{\tilde{U} \to \infty}  \tilde{U} \left( d^\dag d - 1 \right)^2 \biggr\}.
\eea
which is identical to that for the FO fixed point with the addition of an infinite $U$ Coulomb repulsion on the dot level.  The corresponding spectrum of the LM fixed point will be the same as that for the FO fixed point with the exclusion of all of those states for which the dot level is empty or doubly-occupied as these now have an infinite energy cost.  The lowest energy levels of the LM fixed point are listed in Table~\ref{table:LMspec}.
\begin{table}
\begin{center}
\begin{tabular}{|cccc|cccc|}
\hline
Energy & Num.~Value & $Q$ & $2S$ & Energy & Num.~Value & $Q$ & $2S$ \\
\hline
0 & 0.0000 & 0 & 1 & $2\eta_{1e}^+ + \eta_{1o}^-$ & 0.4485 & 1 & 0 \\
\cline{1-4}
$\eta_{1e}^+$ & 0.1495 & 1 & 0 & & & 1 & 2 \\
& & 1 & 2 & $\eta_{1e}^+ + 2\eta_{1o}^-$ & & -1 & 0 \\
$\eta_{1o}^-$ & & -1 & 0 & & & -1 & 2 \\
\cline{5-8}
& & -1 & 2 & $2\eta_{1e}^+ + 2\eta_{1o}^-$ & 0.5980 & 0 & 1\\
\hline
$2 \eta_{1e}^+$ & 0.2990 & 2 & 1 & $\eta_{1o}^+$ & 1.3580 & 1 & 0 \\
$2 \eta_{1o}^-$ & & -2 & 1 & & & 1 & 2 \\
$\eta_{1e}^+ + \eta_{1o}^-$ & & 0 & 1 &  $\eta_{1e}^-$ & & -1 & 0 \\
& & 0 & 1 & & & -1 & 2 \\
& & 0 & 3 & & & & \\
\hline
\end{tabular}
\end{center}
\caption{\label{table:LMspec} The lowest energies and associated total charge $Q$ and total spin $S$ quantum numbers of the local moment (LM) NRG fixed point of the Aharanov-Bohm ring model for $N$ odd.  The single-particle energy levels are the same as in Table~\ref{table:FOspec} using $\tilde{V}_p = 3.0$ and $\Lambda = 2.5$.   All energies within a particular box are equal by eq.~(\ref{spEnergyRelation}).  }
\end{table}

To determine the spectrum of the SC fixed point, we must first identify the linear combination of electrons that screens the local moment on the quantum dot.  However, as discussed, we do not transform to scattering states in the NRG and so we simply use the combination in eq.~(\ref{HcontNRG}) that couples directly to the quantum dot as the screening channel, keeping the potential scattering terms in the Hamiltonian, allowing the numerics to account for those terms directly.  That is, we transform the original Hamiltonian, eq.~(\ref{HcontNRG}), by rotating to a basis
\bea
\psi_{1k} & = & \cos \frac{\vphi}{2} e_k + \sin \frac{\vphi}{2} o_k \\
\psi_{2k} & = & \sin \frac{\vphi}{2} e_k - \cos \frac{\vphi}{2} o_k
\eea
so that
\begin{widetext}
\bea
\nonumber H & = & v_F \sum_{b=1,2} \int dk \, k \, \psi_{bk}^\dag \psi_{bk} + V_d \int dk \left( \psi_{1k}^\dag d + \hc \right) + \frac{U}{2} \left( d^\dag d - 1 \right)^2 \\
& & - V_p \int dk \, dk^\p \left[ \cos \vphi \left( \psi_{1k}^\dag \psi_{1 k^\p} - \psi_{2k}^\dag \psi_{2 k^\p} \right) + \sin \vphi \left( \psi_{1k}^\dag \psi_{2k^\p} + \hc \right) \right]
\eea
and take $\psi_{1k}$ as the screening channel.  

The strong-coupling fixed point involves the $\psi_{1}(x=0)$ electrons forming a singlet with the dot local moment, effectively removing the $\psi_1(0)$ and $d$ degrees of freedom from the dynamics and giving rise to a $\pi/2$ phase shift in the $\psi_1$ channel.  One can then apply the standard NRG transformations and approximations to the resulting model in order to obtain a Wilson chain NRG form of the SC fixed point Hamiltonian.  The $\pi/2$ phase shift is implemented by shrinking the length of the $\psi_1$ Wilson chain by one site representing the removal of the site that is entangled in the Kondo singlet.  

The result is
\bea
\nonumber H_{N, SC} & =  \Lambda^{(N-1)/2} \biggl\{ &  \sum_{n=0}^{N-2} \Lambda^{-(n+1)/2} \xi_n \left( f_{n,1}^\dag f_{n+1,1} + \hc \right) 
+ \sum_{n=0}^{N-1} \Lambda^{-\frac{n}{2}} \xi_n \left( f_{n,2}^\dag f_{n+1,2} + \hc \right) \\
\nonumber & & - \Lambda^{-\frac{1}{2}} \tilde{V}_{p}^\prime \cos \vphi f_{0,1}^\dag f_{0,1} +\tilde{V}_{p} \cos \vphi f_{0,2}^\dag f_{0,2} \\
\label{HNSC} & &- \Lambda^{- \frac{1}{4}} \sqrt{\tilde{V}_{p}^\prime \tilde{V}_{p}} \sin \vphi \left( f_{0,1}^\dag f_{0,2} + \hc \right) \biggr\}.
\eea
\end{widetext}
Here, $f_{n,1}$ and $f_{n,2}$ are the NRG Wilson chain operators derived from $\psi_{1}$ and $\psi_{2}$ respectively.  The differing $\Lambda$ prefactors are due to the normalizations required for the two different length chains.  We have also added an additional factor, $\tilde{V}_p^\prime$, which arises from the additional potential scattering term in the screening channel discussed in section~\ref{sec:potscat}.  For now we simply take it as a single fitting parameter and return to its precise analysis in \S~\ref{sec:nrgVR}.  Since the $\psi_2$ channel does not participate in the screening of the quantum dot, we do not expect any additional potential scattering term proportional to $f_{0,2}^\dag f_{0,2}$.  For the cross-term involving $f_{0,1}^\dag f_{0,2} + \hc$, we simply take the geometric mean of the two potential scattering terms of the two channels and find that this provides a good fit to the NRG data. 

To obtain the SC fixed point spectrum, we first find the single-particle energy levels by numerically diagonalizing eq.~(\ref{HNSC}) for a finite value of $N$.  As before, we find that for $N > 10$ (approximately), the energy levels depend only on the parity of $N$ and not on its precise value.  Unlike the FO and LM fixed point spectra, the resulting energy levels will depend on the flux $\vphi$ in addition to the $\tilde{V}_p$ dependence.  Similar to eq.~(\ref{HNFOdiag}), we can write the SC fixed point Hamiltonian in terms of the single particle and hole excitations
\begin{widetext}
\be
\label{HNSCdiag} H_{N, \mathrm{SC}} = \left\{ 
\begin{array}{ll}
\sum_{n=1}^{(2N+1)/2} \left( \nu^+_{n}(\tilde{V}_p, \tilde{V}_p^\prime, \vphi) g_{n}^\dag g_{n} + \nu^-_{n}(\tilde{V}_p, \tilde{V}_p^\prime, \vphi) h_{n}^\dag h_{n} \right) & \quad, N \, \textrm{odd} \\
\sum_{n=1}^{(2N+1)/2} \left( \hat{\nu}^+_{n}(\tilde{V}_p, \tilde{V}_p^\prime, \vphi) g_{n}^\dag g_{n} + \hat{\nu}^-_{n}(\tilde{V}_p, \tilde{V}_p^\prime, \vphi) h_{n}^\dag h_{n} \right) & \quad, N \, \textrm{even}.
\end{array}
\right.
\ee 
\end{widetext}
Because of the coupling of the 1 and 2 channels in eq.~(\ref{HNSC}), the quasiparticle excitations cannot be labelled by a channel index since it is no longer a good quantum number.

The full many-body spectrum is constructed by combining these single-particle excitations in such a way as to respect Fermi statistics.  The effect of the Kondo singlet, in addition to the $\pi/2$ phase shift already implemented in eq.~(\ref{HNSC}), is simply to add an additional charge to the quantum numbers of the quasiparticle excitations due to the Fermion doing the screening.   The lowest such energies are listed in Table~\ref{table:SCspec}.

\begin{table}
\begin{center}
\begin{tabular}{|ccccc|ccccc|}
\hline
& \multicolumn{2}{c}{Num. Value} & & & & \multicolumn{2}{c}{Num. Value} & & \\
Energy & $H_{N,SC}$ & NRG & $Q$ & $2S$ & Energy & $H_{N,SC}$ & NRG & $Q$ & $2S$ \\
\hline
0 & 0.000 & 0.000 & 1 & 0 & $\nu_{2}^- + 2 \nu_{1}^-$ & 0.8153 & 0.8158 & -2 & 1 \\
$\nu_{1}^-$ & 0.0709 & 0.0711 & 0 & 1 & $ \nu_{1}^+$ & 0.8201 & 0.8203 & 2 & 1 \\
$2 \nu_{1}^-$ & 0.1416 & 0.1422 & -1 & 0 & $\nu_{1}^+ + \nu_{1}^-$ & 0.8910 & 0.8914 & 1 & 0 \\
$\nu_{2}^-$ & 0.6737 & 0.6736 & 0 & 1 & & & 0.8914 & 1 & 2 \\
$\nu_{2}^- + \nu_{1}^-$ & 0.7445 & 0.7447 & -1 & 0 & $\nu_{1}^+ + 2\nu_{1}^-$ & 0.9618 & 0.9625 & 0 & 1\\
& & 0.7447 & -1 & 2 & $2\nu_{2}^-$ & 1.3474 & 1.3472 & -1 & 0 \\
\hline
\end{tabular}
\end{center}
\caption{\label{table:SCspec} The lowest energies and associated total charge $Q$ and total spin $S$ quantum numbers of the strong coupling (SC) NRG fixed point of the Aharanov-Bohm ring model for odd $N$.  The NRG parameters used are $\tilde{V}_p = 3.0$ and $\vphi = 1.047$.  The same parameters were used to determine the energy levels of $H_{N,SC}$ where a value of $\tilde{V}_p^\prime = 2.885$ was found to reproduce the NRG data. }
\end{table}

Guided by the results of the transformations of section~\ref{sec:model}, we have now identified the three fixed points of the Aharanov-Bohm quantum dot model and written the corresponding Hamiltonians in a Wilson chain form, eqs.~(\ref{HNFO}), (\ref{HNLM}), and~(\ref{HNSC}).  This allows us to determine the fixed point spectra, the lowest values of which have been listed in Tables~\ref{table:FOspec}, \ref{table:LMspec}, and~\ref{table:SCspec}.  We are now prepared to test these predictions by comparing these spectra with the actual energy levels that are computed in the NRG.   

\begin{figure}
\begin{center}

\psfrag{n1}{}
\psfrag{n2}{}
\psfrag{n3}{}
\psfrag{n4}{}
\psfrag{n5}{}
\psfrag{n6}{}
\psfrag{n7}{}
\psfrag{n8}{}

\psfrag{m1}{}
\psfrag{m2}{}
\psfrag{m3}{}
\psfrag{m4}{}
\psfrag{m5}{}
\psfrag{m6}{}
\psfrag{m7}{}

\includegraphics[width=0.45\textwidth, clip=true]{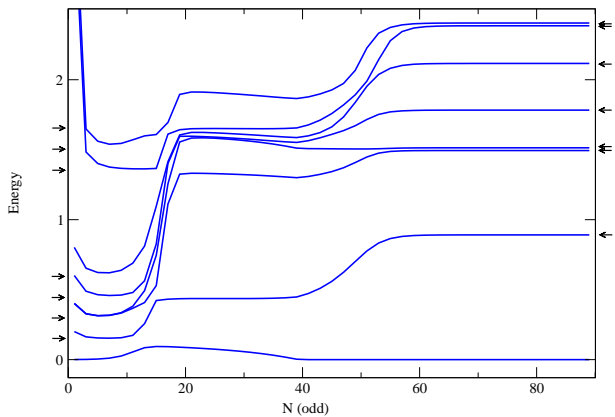}
\end{center}
\caption{\label{fig:energies-fixedpoints}  (Color online) The lowest energy levels with quantum numbers $Q=1$, $S=0$ as produced by the NRG as a function of odd $N$.  The values of the predicted fixed point energies from Tables~\ref{table:FOspec} and \ref{table:LMspec} are indicated by arrows on the left side and energies from Table~\ref{table:SCspec} are indicated on the right.  The parameters used to generate this plot are $\Gamma/D = 0.0003142$, $U/D = 0.001$, $\nu V_p = 1.05$, and $\vphi = 1.047$.  Here we see the unstable FO fixed point is approached for $5 < N < 10$, the unstable LM fixed point for $19 < N < 33$, and the stable SC fixed point for $N>60$.}
\end{figure}

This comparison is achieved by looking at the flow of the energy levels of each $H_N$ (as defined in eq.~(\ref{HN})) for increasing $N$.   An example is shown in Fig.~\ref{fig:energies-fixedpoints} where we have plotted the lowest few energy levels of the $Q=1$, $S=0$ subspace as a function of odd $N$.   It is shown that the fixed point spectra predicted above are indeed approached in the appropriate regime.  For example, for $5 < N < 10$, all of the energies of the $Q=1$, $S=0$ subspace of the unstable FO fixed point are approached with the proper numerical value as given in Table~\ref{table:FOspec}.  Similarly, for $19 < N < 33$, the predicted energy levels of the LM fixed point (Table~\ref{table:LMspec}) are approached.  The same is true for the SC fixed point where, in Table~\ref{table:SCspec}, $\tilde{V}_p^\prime$ is fit in order to produce the fixed point spectrum produced by the NRG algorithm (for the parameters used to generate the NRG data, a value of $\tilde{V}_p^\prime = 2.885$ was found to give the best fit).

\begin{figure}
\begin{center}
\psfrag{p0}{$0$}
\psfrag{p111111}{$\pi/6$}
\psfrag{p2}{$\pi/3$}
\psfrag{p3}{$\pi/2$}
\includegraphics[width=0.45\textwidth, clip=true]{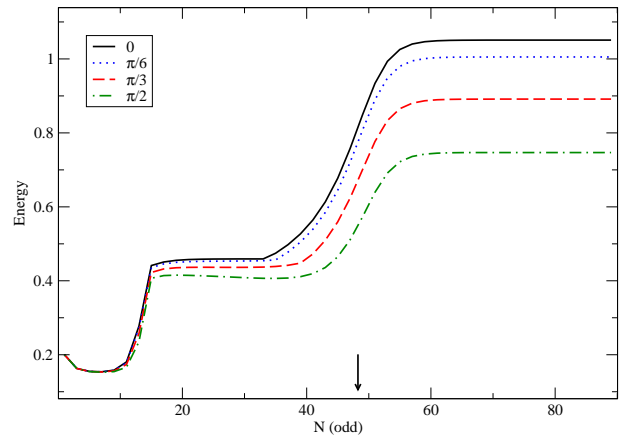}
\end{center}
\caption{\label{fig:energies_multiphi} (Color online) The lowest non-zero energy level with quantum numbers $Q=1$, $S=0$ as produced by the NRG as a function of odd $N$.  The different lines correspond to different values of the flux $\vphi$.  The parameters used to generate this plot are the same as in Fig.~\ref{fig:energies-fixedpoints}.  The value of $N_K$, related to the Kondo temperature via eq.~(\ref{TKN}), is indicated by the arrow and is the same for all values of $\vphi$. }
\end{figure}

In Fig.~\ref{fig:energies_multiphi}, we show a similar plot of a single energy level as a function of odd $N$ in the $Q=1$, $S=0$ subspace where the different lines indicate energies produced from different values of the flux $\vphi$.  Here we see that, as predicted, the FO and LM fixed point energy levels that are approached are independent of $\vphi$ whereas those of the SC fixed point are strongly flux dependent.   The slight flux dependence that appears in the LM region is probably due to the fact that $\tilde{V}_d$ is not quite zero (\textit{i.e.}\ the LM fixed point is approached but never reached).  Indeed, the flux dependence of the energy levels in this region decreases the closer the LM fixed point is approached.

\begin{figure}
\begin{center}
\includegraphics[width=0.45\textwidth, clip=true]{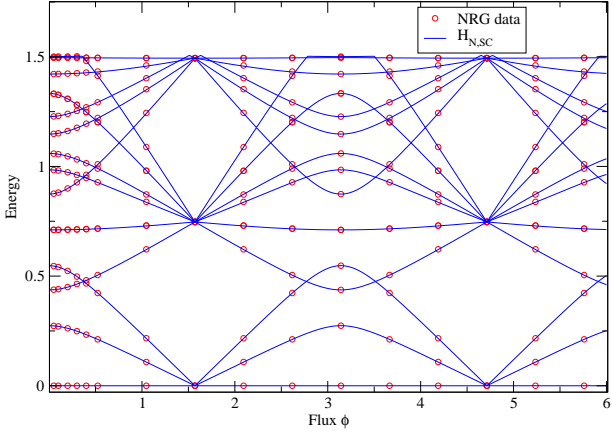}
\end{center}
\caption{\label{fig:SCfixedpoint-multiphi} (Color online) The lowest NRG energy levels of the final stable fixed point (red circles) are plotted as a function of flux $\vphi$ and compared with those given by the fixed point Hamiltonian of eq.~(\ref{HNSC}) (solid blue lines).  The parameters used to generate this plot are the same as in Fig.~\ref{fig:energies-fixedpoints} except here we use a value of $\nu V_p = 0.525$.  The single value of $\tilde{V}_p^\prime$ was tuned in order to fit the NRG fixed point energy levels for $\vphi = 0$.  This single parameter is able to reproduce the predicted flux dependence.}
\end{figure}

For a more quantitative analysis of this flux dependence, we plot the lowest NRG energy levels of the final, stable fixed point with those predicted by diagonalizing the Hamiltonian of eq.~(\ref{HNSC}) as a function of $\vphi$ in Fig.~\ref{fig:SCfixedpoint-multiphi}.  The fact that a single parameter fit of $\tilde{V}_p^\prime$ perfectly reproduces the flux dependence of the entire NRG fixed point spectrum strongly supports the validity of the above RG analysis.  Indeed, because the SC fixed point is stable, we can explicitly compare the fixed point spectrum produced by the NRG with that predicted by eq.~(\ref{HNSC}) as we have done in Table~\ref{table:SCspec} for the first few levels.

\subsubsection{Kondo Temperature from the NRG}
\label{sec:TKNRG}
In \S~\ref{sec:TK}, we derived an expression for the Kondo temperature in terms of the inter-lead tunneling $t^\p$, the flux $\vphi$, and the Fermi momentum $k_F$, eq.~(\ref{lnTK}).  For the particle-hole symmetric value of $k_F = \pi/2a$, this expression takes the simple form of eq.~(\ref{TKvstp}).  It is this latter form that can be compared to the NRG which was derived from a model with a particle-hole symmetric linear dispersion.  

To do this, we must write eq.~(\ref{TKvstp}) in terms of the Anderson model parameters appearing in the NRG Wilson chain form of the Hamiltonian, eq.~(\ref{HNRGinf}), that serve as input to the NRG.   First, we define an effective Kondo coupling for the continuum model of eq.~(\ref{HcontNRG})
\be
J = 2 \frac{4 V_d^2}{U}.
\ee
The factor of 2 is included because eq.~(\ref{HcontNRG}) involves coupling to both the even and odd channels whereas, in eq.~(\ref{Jdef}), we defined $J$ for the screening channel only.  In transforming to the screening channel, a factor of $\sqrt{2}$ appears in $\tilde{V}_{dk}$ resulting in the derived $J$ acquiring a factor of 2 which we account for here explicitly so that we can compare the NRG results with those derived analytically. 

Using the definition of $\Gamma$,~(\ref{Gammadef}), we get 
\be
\label{nuJTK}
\nu J = 4 \Gamma / (\pi U).
\ee  
Next, we recall that $V_p = - v_{k_F k_F} = 2 a t \tau^\p / \pi$ so that we can write $\tau^\p = \pi \nu V_p$.  The resulting expression is
\be
\label{TKvsVP_NRG}
\ln \frac{T_K}{T_K^0} = - \frac{\pi^2 (\nu V_p)^2}{2 \nu J}.
\ee
The right hand side of this equation, together with eq.~(\ref{nuJTK}), now contains parameters related directly to the input parameters of the NRG.

We now must extract the Kondo temperature from the NRG data for multiple values of $V_p$ and $J$ in order to confirm the validity of~(\ref{TKvsVP_NRG}).  The Kondo temperature is defined as the energy scale at which the screening of the local moment takes place and the Hamiltonian crosses over to the stable SC fixed point.  In the NRG, $T_K$ will be related to the value of $N$ at which the energy levels cross over from that of the LM or FO fixed point to those of the SC as described in the previous section.  This value of $N$, which we denote $N_K$, at which the crossover takes place can be related to a corresponding energy scale using eq.~(\ref{EN}), namely
\be
\label{TKN}
k_B T_K \approx \frac{1}{2} \left(1 + \Lambda^{-1}\right) \Lambda^{- (N_K-1)/2} D.
\ee
One simply has to extract the value of $N_K$ from the NRG energy level data in order to obtain $T_K$.  In practice, we measure $N_K$ for the lowest 20 NRG energy levels and use the mean value $\langle N_K \rangle$ to determine $T_K$.

\begin{figure}
\begin{center}
\includegraphics[width=0.45\textwidth, clip=true]{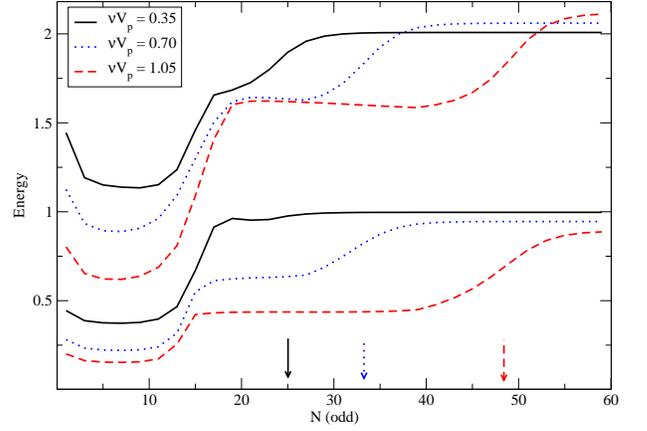}
\end{center}
\caption{\label{fig:energies_multiVp}  (Color online) Select energy levels with quantum numbers $Q=1$, $S=0$ as produced by the NRG as a function of odd $N$.  The different lines correspond to different values of the inter-lead coupling $V_p$ and the arrows of the same line type indicate the approximate value of $N_K$ at which the Kondo crossover takes place for each case.   The parameters used to generate this plot are the same as in Fig.~\ref{fig:energies-fixedpoints}.}
\end{figure}

In Figure~\ref{fig:energies_multiVp}, we have plotted select NRG energy levels as a function of $N$ for different values of $V_p$.  There is clearly a trend of increasing $N_K$ with increasing $V_p$ which, from eq.~(\ref{TKN}), indicates a decrease in $T_K$ as a function of $V_p$ as predicted in eq.~(\ref{TKvsVP_NRG}).  Furthermore, if one looks at Figure~\ref{fig:energies_multiphi},  there is clearly no change in the value of $N_K$ for the different values of flux $\vphi$ indicating that there truly is no flux dependence in $T_K$ when $k_F = \pi / (2a)$.

\begin{figure}
\begin{center}
\includegraphics[width=0.45\textwidth, clip=true]{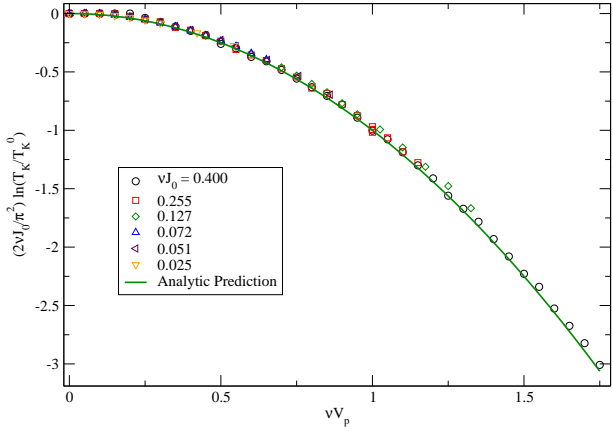}
\end{center}
\caption{\label{fig:TKvsVp}  (Color online) A plot of $T_K$ (symbols) as extracted from the NRG as a function of the input value of $\nu V_p$.  The different symbols describe data with different input parameters giving rise to different effective Kondo couplings.  The solid line indicates the prediction described in eq.~(\ref{TKvsVP_NRG}).}
\end{figure}

For a more quantitive comparison, we have plotted the value of $T_K$ extracted from the NRG as a function of $\nu V_p$ in Figure~\ref{fig:TKvsVp} for multiple values of the Kondo coupling $J$.  The analytic form predicted in eq.~(\ref{TKvsVP_NRG}) provides an excellent, parameter-free fit to the numerical data. 

\subsection{Phase Shifts and $V_R$}
\label{sec:nrgVR}
As discussed in \S~\ref{sec:conductance}, the SC fixed point is comprised of two independent Fermi liquids characterized by two phase shifts.  These phase shifts are determined by the eigenvalues of the S-matrix of eq.~(\ref{Smatrixeo}).  In this section, we wish to compare these two predicted phase shifts with those derived from the NRG.  

Once again, given the particle-hole symmetric formulation of the NRG, we can only make this comparison at the special value of $k_F = \pi / (2a)$.  In this special limit, one can see from eq.~(\ref{deltadef}) that $\delta^+ = - \delta^- \equiv \delta$ where $\tan \delta = \tau^\p$.  We further simplify to the symmetric case $t_{d-} = t_{d+} = t_d$.  In this case, the two eigenvalues of the $S$-matrix are 
\be
\lambda_\pm = - e^{i \delta_R} \left(i A \pm \sqrt{1 - A^2} \right)
\ee
with
\be
\label{psA}
A \equiv \cos 2 \delta \sin \delta_R + \sin 2 \delta \cos \delta_R \cos \vphi.
\ee
Writing these as pure phases $\lambda_\pm = e^{2 i \alpha_\pm}$, the phase shifts are given by
\be
\label{phase_shifts_full}
\cos 2 \alpha_\pm = A \sin \delta_R \mp \sqrt{1 - A^2} \cos \delta_R.
\ee
In the special case of $\vphi = 0$ when the two channels fundamentally decouple, one obtains
\be
\cos 2 \alpha_\pm = \mp \cos( 2 \delta + \delta_R \pm \delta_R), \qquad (\vphi = 0)
\ee
or
\bea
\label{deltaPlus_phiZero}
\alpha_+ & = & \frac{\pi}{2} - \delta - \delta_R, \qquad (\vphi = 0) \\
\label{deltaMinus_phiZero}
\alpha_- & = & \delta, \qquad \qquad \qquad (\vphi = 0).
\eea
The two phase shifts $\alpha_\pm$ fully define the strong-coupling fixed point spectrum.

\subsubsection{Phase shifts from the NRG}
First, we consider a system of two independent Fermi liquids on a finite line of length $L$ and with linear dispersion relations.  The energy levels will then take the form
\be
\label{linearspectrum}
\eps_n^{(i)} = \frac{2 \pi v_F}{L} \left(q - \frac{\delta_i}{\pi} \right)
\ee
where $q \in \mathbb{Z}$ and $\delta_i$ are the phase shifts in the $i^{\mathrm{th}}$ channel.     

The situation with the NRG is not quite so simple due to the non-uniform hopping in the Wilson chain that goes like $\Lambda^{-n/2}$ at the $n^{\mathrm{th}}$ site.  However, one can still extract a sensible phase shift describing the overall shift of the (non-uniform) energy spectrum.  We present a method for extracting these phase shifts from the NRG data that is similar to that used in~\cite{hofstetter2004} though we are much more modest about the claimed analogy between the non-uniform NRG spectrum and that of eq.~(\ref{linearspectrum}).

As discussed in \S~\ref{sec:FPABmodel}, the many-body spectrum of the strong-coupling fixed point is built up of two channels of single-particle excitations, both of which we denoted together as $\nu_n^\pm$  where  the $\pm$ superscripts indicate whether the excitation is that of a particle (+) or a hole (-).  With knowledge of only the  total charge $Q$ and total spin $S$ quantum numbers of each many-body energy from the NRG, one can identify the single-particle energy levels for each of the two channels which we denote as $\nu^\pm_{n+}$ and $\nu^\pm_{n-}$.  It is from these that we estimate the phase shift in each channel.

\begin{figure}
\begin{center}
\psfrag{nmm}{$\nu_{1-}^-$}
\psfrag{nmp}{$\nu_{1-}^+$}
\psfrag{npm}{$\nu_{1+}^-$}
\psfrag{npp}{$\nu_{1+}^+$}
\psfrag{pc}{+ channel}
\psfrag{mc}{- channel}
\includegraphics[width=0.45\textwidth, clip=true]{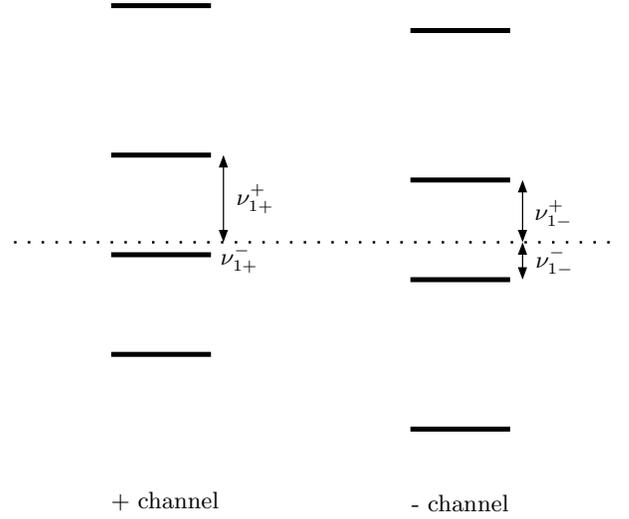}
\end{center}
\caption{\label{fig:NRGphaseshift}  Energy level diagrams of the single-particle NRG energy levels of the two channels.  The shift of each relative to the Fermi energy (here indicated by the dotted line) defines the phase shift in each channel.}
\end{figure}
For clarity, let us assume that the NRG chain length $N$ is even~\cite{footnote1}  and that the four lowest energy levels are ordered such that $\nu_{1+}^- < \nu_{1-}^- < \nu_{1-}^+ < \nu_{1+}^+$, as depicted in figure~\ref{fig:NRGphaseshift}.  
The phase shift in each of the channels is going to be proportional to the lowest single-particle energy level in each channel, in this case, $\nu_{1+}^-$ and $\nu_{1-}^-$.  However, because of the non-uniform $\Lambda$-dependent spacing of the energy levels, we normalize each phase shift by the lowest energy level spacing in their respective channels.  That is, we define the phase shift as
\bea
\alpha_+ & = & \frac{\nu_{1+}^-}{\nu_{1+}^- + \nu_{1+}^+} \pi \\
\alpha_- & = &  \frac{\nu_{1-}^-}{\nu_{1-}^- + \nu_{1-}^+} \pi.
\eea
If the channels are shifted in the other direction relative to the Fermi energy (that is, if the lowest single-particle excitation is that of a particle instead of a hole: $\nu_{1+}^+ < \nu_{1-}^+ < \nu_{1-}^- < \nu_{1+}^-$), the phase shifts are taken to be
\bea
\alpha_+ & = & - \frac{\nu_{1+}^+}{\nu_{1+}^- + \nu_{1+}^+} \pi \\
\alpha_- & = &  - \frac{\nu_{1-}^+}{\nu_{1-}^- + \nu_{1-}^+} \pi.
\eea

One can now extract the values of $\nu_{1i}^\pm$ from the many-body NRG energy spectrum obtained by diagonalizing $H_N$ as described in \S~\ref{sec:FPABmodel}.  Assume that $N$ is sufficiently high such that the RG has reached the strong-coupling fixed point.  The ground state, describing no particles or holes and set arbitrarily to $E_0=0$, will have total spin quantum number $S=0$ and a charge quantum number of $Q_0 = +1$ or $Q_0 = -1$ depending on whether the lowest single-particle energy is a hole or a particle respectively.  Let us assume that $Q_0 = +1$ for clarity.  Then, the values of $\nu_{1+}^-$ and $\nu_{1-}^-$ are given by the two lowest many-body energies with a charge quantum number of $Q=0$ and spin quantum number $S=1/2$ (these lowest energies would be $\nu_{1+}^+$ and $\nu_{1-}^+$ if $Q_0 = -1$).  The values of $\nu_{1+}^+$ and $\nu_{1-}^+$ are given by the lowest many-body energies with charge quantum number $Q=+2$ and spin quantum number $S=1/2$ (in the case of $Q_0 = -1$, $\nu_{1+}^-$ and $\nu_{1-}^-$ would be given by the lowest $Q=-2$, $S=1/2$ many-body energies).    In this way, one can extract the single-particle/hole energies and estimate the phase shifts from the NRG data.

\begin{figure}
\begin{center}
\includegraphics[width=0.45\textwidth, clip=true]{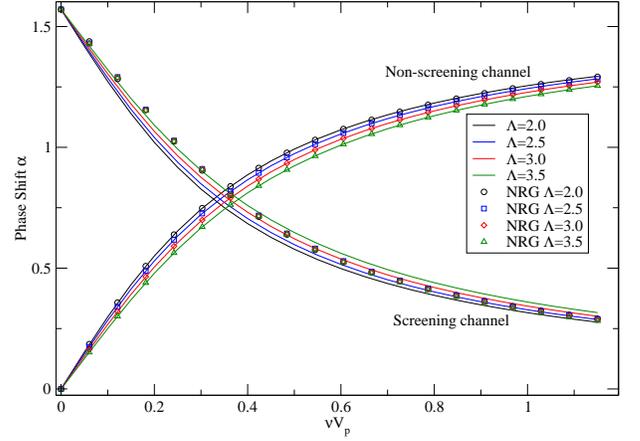}
\end{center}
\caption{\label{fig:ps_LambdaDependence} (Color online) The phase shift of the two channels as a function of $V_p = -v_{k_F k_F}$ from the NRG data with an effective Kondo coupling $\nu J = 0.255$ and zero flux.  In this case, the two channels are independent with the screening channel phase shift starting at $\pi / 2$ for $\nu V_p = 0$ and the non-screening channel phase shift starting at $0$ for $\nu V_p = 0$.  The symbols are the phase shifts obtained from the full NRG many-body energy levels.  The ascending solid lines are the phase shifts obtained from the single-particle energy levels of a single non-interacting Wilson chain with potential scattering $V_p$ and the descending solid lines are $\pi/2$ minus the ascending lines.  The solid lines do not take into account the small correction due to the additional potential scattering $V_R$ that occurs in the screening channel.} 
\end{figure}

As an illustration of the $\Lambda$ dependence of these phase shifts, we consider the simple case of zero-flux, $\vphi = 0$.  In this case, the original Hamiltonian can be completely decoupled into two separate channels and so the two channels operate completely independently.  The channel coupled to the quantum dot is the screening channel and so obtains a $\pi / 2$ phase shift in addition to that given by $\delta$ whereas the other channel is non-interacting with only a potential scattering phase shift $-\delta$.  This can be seen clearly in figure~\ref{fig:ps_LambdaDependence} where we have plotted the two phase shifts as a function of $V_p$ (recall from \S~\ref{sec:FPABmodel} that $V_p = -v_{k_F k_F}$ and so is related to $t^\p$ via eq.~(\ref{vkkp})).

The most striking feature of figure~\ref{fig:ps_LambdaDependence} is the different $\Lambda$ dependence in the phase shift of the screening and non-screening channels obtained from the NRG data.  To help understand this, we have plotted as solid lines the phase shifts that one would expect in a non-screening and screening Wilson chain (we ignore the effects of the small correction due to $V_R$ for now).  For the non-screening channel, one can diagonalize directly the Wilson chain Hamiltonian with a potential scattering $V_p$ at the first site using different values of $\Lambda$ and so obtain the single-particle energy spectra directly without having to perform the NRG.  From this direct single-particle spectra one can define the phase shift as described above and these are plotted as the solid ascending lines.  As can be seen, these match perfectly the phase shifts in the non-screening channel obtained from the NRG data, as they must.  

To leading order (again, neglecting $V_R$), one might expect the phase shift in the screening channel to be simply $\pi / 2$ minus the above $\Lambda$-dependent phase shifts since the potential scattering in the screening channel is equal in magnitude but opposite in sign to that in the non-screening channel.      We have plotted this expectation as the descending solid lines in the figure.  On the contrary, the phase shifts obtained directly from the NRG data show very little $\Lambda$ dependence compared with the non-screening channel.  The precise reason for this is unknown though it may be due to a similar $\Lambda$ dependence in the additional potential scattering $V_R$ as shown in figure~\ref{fig:VR_LambdaDependence} that is compensating for the $\Lambda$ dependence of the bare $\delta$ phase shift.  Despite this, we nevertheless obtain good support for our prediction of the phase shifts from the tight-binding model. 

In the remainder of our analysis, we will use the $\Lambda$ dependent phase shift obtained from diagonalizing the potential scattering Wilson chain discussed above for the bare phase shift $\delta$ generated by $V_p$ that appears in eqs.~(\ref{psA}), (\ref{deltaPlus_phiZero}), and~(\ref{deltaMinus_phiZero}).  See~\cite{krishnamurthy1980b} for more information on the $V_p$ dependence of the NRG spectrum.

\subsubsection{NRG evidence for $V_R$}
\label{NRGVR}
We now turn our attention back to the additional potential scattering $V_R$ that was derived in section~\ref{sec:potscat}.  Having shown that the phase shifts can be extracted from the NRG, we can now compare the predicted phase shifts in eq.~(\ref{phase_shifts_full}) with those of the NRG.  For simplicity, we continue to assume $t_{d-} = t_{d+} = t_d$ and $\eps_d = -U/2$.  To compare our analytic results with those of the NRG, we use the same correspondence as was used in section~\ref{sec:TKNRG}, namely $\tau^\p = \pi \nu V_p$ and $\nu J = 4 \Gamma / (\pi U)$.

\begin{figure}
\begin{center}
\includegraphics[width=0.45\textwidth, clip=true]{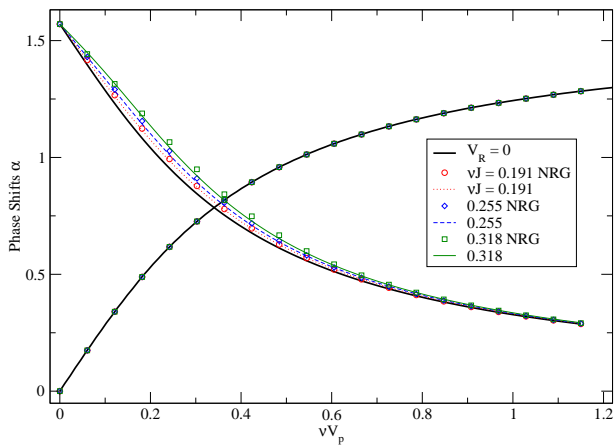}
\end{center}
\caption{\label{fig:deltavsVp} (Color online) The phase shifts of the two channels as a function of $V_p$.  The symbols denote phase shifts derived from the NRG data while the lines are the analytic predictions.  The solid black line is the curve expected if there is no additional potential scattering (\textit{i.e.}\ $V_R = 0$).  We have set $\vphi = 0$ to generate this plot.} 
\end{figure}
We focus first on the case of zero flux, $\vphi = 0$, where the phase shifts take an especially simple form given in eqs.~(\ref{deltaPlus_phiZero}) and~(\ref{deltaMinus_phiZero}).  These two phase shifts are plotted in figure~\ref{fig:deltavsVp} as a function of $V_p$ where the symbols indicate those values derived from the NRG data while the lines are the analytic prediction from the tight-binding model.  Here we see that, indeed, only the phase shift of the screening channel (the one that obtains $\pi / 2$ when $V_p = 0$) deviates from the $V_R = 0$ prediction, indicating that an additional phase shift is generated in the screening channel only.  However, $V_R$ provides only a small correction so it is easier to extract $V_R$ from the NRG phase shifts and compare its functional form directly with that of eq.~(\ref{VR}). 

To extract $V_R$, we take the $\arctan$ of the derived NRG phase shift and subtract from that the $\pi/2$ contribution arising from the Kondo screening as well as the bare phase shift $\delta$ due to $V_p$.  This latter phase shift will be $\Lambda$ dependent and can be calculated numerically as described in~\cite{krishnamurthy1980b}.

\begin{figure}
\begin{center}
\includegraphics[width=0.45\textwidth, clip=true]{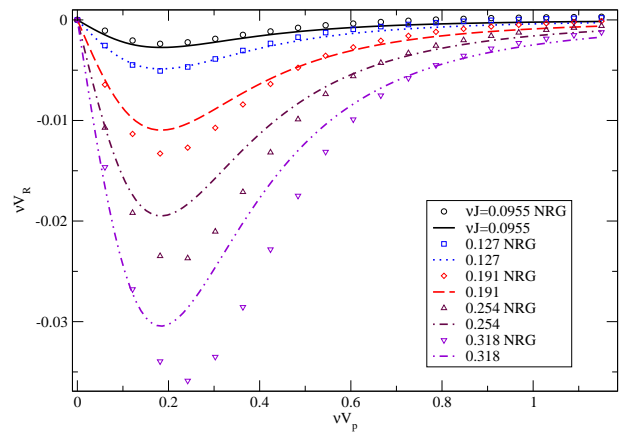}
\end{center}
\caption{\label{fig:VRvsVp} (Color online) The additional potential scattering $V_R$ as derived from the NRG phase shifts (symbols) and from the analytic tight-binding model (lines) for various values of the effective Kondo coupling $J$.  We have set $\vphi = 0$ to generate this plot.} 
\end{figure}
In figure~\ref{fig:VRvsVp}, we compare directly the predicted dependence of $V_R$ on $V_p$ with that derived from the NRG phase shifts for various values of $J$.   We find that both analytic and numeric calculations of $V_R$ share the same qualitative behaviour, peaking around $\nu V_p \approx 0.3$ (corresponding to $t^\p \approx t$ in the original tight-binding model), but that precise quantitative agreement is not obtained.   The nature of this disagreement is discussed in section~\ref{sec:discussion}.

\begin{figure}
\begin{center}
\includegraphics[width=0.45\textwidth, clip=true]{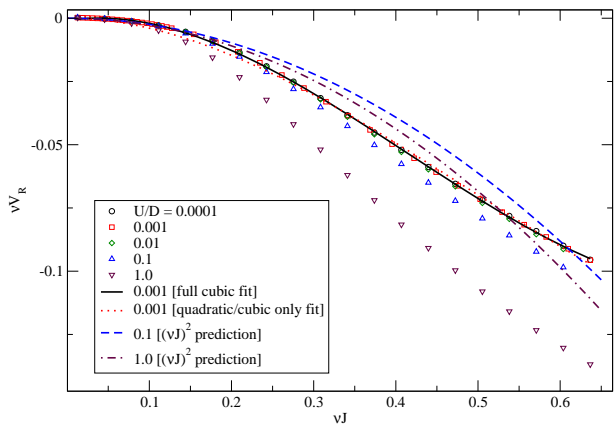}
\end{center}
\caption{\label{fig:VRvsJ} (Color online) The additional potential scattering as determined from the NRG phase shifts as a function of effective Kondo coupling $J$.  Each different symbol uses a fixed value of the dot Coulomb repulsion $U$ while varying $\Gamma$ such that the range of $J$, given in eq.~(\ref{nuJTK}), is roughly the same for each iteration.  The solid line presents the best fit third degree polynomial to the $U/D =  0.001$ points.  In this data, $\nu V_p = 0.3$ and $\vphi = 0$. } 
\end{figure}
We next turn our attention to testing the $(\nu J)^2$ dependence in eq.~(\ref{VR}) by plotting the value of $V_R$ as determined from the NRG phase shifts versus $\nu J$ in figure~\ref{fig:VRvsJ}.  The most striking characteristic is the apparent deviation from universal behaviour as $U/D$ approaches unity.  We see that this trend is captured by the $U$ dependence in eq.~(\ref{VR}) but that precise quantitative agreement is elusive, perhaps because of the presence of a cubic term which we do not consider.  A complete analysis of $V_R$ with an Anderson impurity rather than reducing, via the Schrieffer-Wolff transformation, to one with a spin impurity may elucidate the nature of this behaviour.

For further analysis, we fit the largest data set with $U/D = 0.001$ to a third degree polynomial of the form
\be
\label{cubicfitfull}
V_R = a_0 + a_1 (\nu J) + a_2 (\nu J)^2 + a_3 (\nu J)^3.
\ee
A third degree polynomial was chosen instead of a second degree function because the data goes to quite large values of $\nu J$ where we expect our second order analysis to break down.  The values of the parameters are tabulated in Table~\ref{table:polyfit}.  It is seen that the coefficients $a_0$ and $a_1$, which we predict to vanish, are indeed at least an order of magnitude lower than the quadratic and cubic coefficients.  Doing another fit neglecting these first two terms, that is, to a form
\be
\label{cubicfit2}
V_R = b_2 (\nu J)^2 + b_3 (\nu J)^3
\ee
gives $b_2 = -0.42$ which is the same order of magnitude as the value of $-0.24$ predicted by eq.~(\ref{VR}).
\begin{table}
\begin{center}
\begin{tabular}{c|c}
\hline
Coefficient & Value \\
\hline
$a_0$ & -0.00081 \\
$a_1$ & 0.045 \\
$a_2$ & -0.62 \\
$a_3$ & 0.50 \\
\hline
$b_2$ & -0.42 \\
$b_3$ & 0.29 \\ 
\hline
\end{tabular}
\end{center}
\caption{\label{table:polyfit} The parameters for the best fit of eqs.~(\ref{cubicfitfull}) and~(\ref{cubicfit2}) to the $U/D = 0.001$ data in figure~\ref{fig:VRvsJ}.}
\end{table}

\begin{figure}
\begin{center}
\includegraphics[width=0.45\textwidth, clip=true]{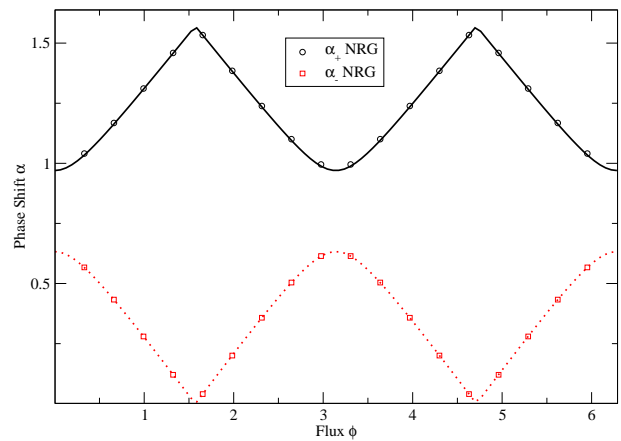}
\end{center}
\caption{\label{fig:psvsphi} (Color online) The phase shifts of the strong-coupling fixed point as determined from the NRG (symbols) and compared with that predicted in eq.~(\ref{phase_shifts_full}) (lines).  Here, the effective Kondo coupling is $\nu J = 0.191$ and $\nu V_p = 0.25$. } 
\end{figure}
Up until this point we have been focussing primarily on the form of the additional potential scattering $V_R$ and so, for simplicity, have taken the flux $\vphi = 0$.  In figure~\ref{fig:psvsphi}, we have plotted the phase shifts $\alpha_\pm$ versus the flux $\vphi$ as derived from the NRG with comparison to the predicted form described in eq.~(\ref{phase_shifts_full}).  There we find the agreement to be quite good and suggests that our predicted flux dependence is robust.  

\begin{figure}
\begin{center}
\includegraphics[width=0.45\textwidth, clip=true]{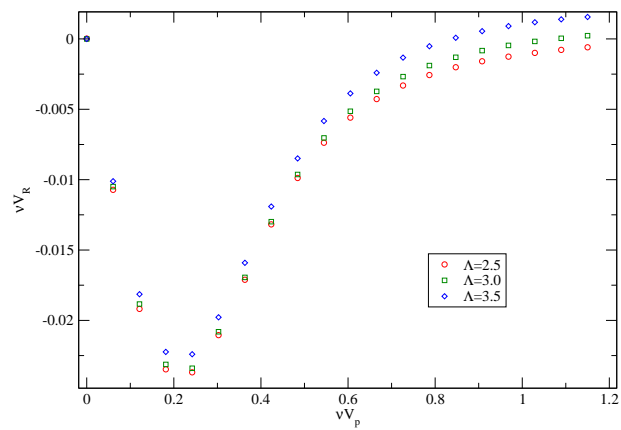}
\end{center}
\caption{\label{fig:VR_LambdaDependence} (Color online) The value of the additional potential scattering $V_R$ as derived from the NRG for various values of $\Lambda$, all using a value of $J = 0.510$.} 
\end{figure}
Finally, we note that, although it seems that the $\Lambda$ dependence of the screening channel phase shift is suppressed (see figure~\ref{fig:ps_LambdaDependence}), there does appear to be some systematic $\Lambda$ dependence in $V_R$ itself as seen in figure~\ref{fig:VR_LambdaDependence}.  In order to take this effect into account, one would need to derive an expression for $V_R$ from the Wilson chain Hamiltonian, eq.~(\ref{HNRGinf}), as opposed to the much simpler tight-binding model as was done in eq.~(\ref{VR}).  We still find convincing agreement of the behaviour of $V_R$ between that derived analytically and from the NRG despite this apparent $\Lambda$ dependence.

\section{Discussion}
\label{sec:discussion}
We have presented a systematic study of a minimal model of an Aharanov-Bohm ring with an embedded quantum dot connected to two conducting leads.  Although aspects of such a system have been studied by other groups in the past~\cite{bulka2001, hofstetter2001, kim2002, aharony2005, simon2005, yoshii2008, davidovich1997, gerland2000, konig2002, konik2004}, our work provides a complete picture of the physics of such a system when the quantum dot chemical potential $\epsilon_d$ is near
$-U/2$ and the system is in the Kondo regime including new effects not discussed previously.  

In particular, we have elucidated precisely how the Kondo effect arises in such a system by identifying the screening channel; we have completely mapped out the renormalization group flow of the system and its dependence on flux $\vphi$ and inter-lead tunneling $t^\p$; we have calculated the dependence of the Kondo temperature and conductance on the same parameters as well as, for the first time, the electron density in the leads (via the factors of $k_F$ appearing throughout); we have calculated the effects of additional potential scattering that arises from the breaking of particle-hole symmetry; we have provided wide numerical support from the NRG for many of our findings that goes beyond simply computing the occupancy of the quantum dot as in~\cite{hofstetter2001} or the dot density of states~\cite{gerland2000}.  Although our work is quantitatively precise, the physical picture that arises has been stated in simple physical terms that fully describes the zero-temperature properties.

It is interesting to compare our results for the Kondo temperature with those of refs.~\onlinecite{simon2005} and~\onlinecite{yoshii2008}.  In the former reference, the authors use a slave boson mean field theory to estimate the Kondo temperature for variable sized rings.  For the smallest configuration with only one site in the ring in addition to the quantum dot, they find a flux dependent Kondo temperature assuming particle-hole symmetric leads, $k_F = \pi / (2a)$.  Although the calculation of ref.~\onlinecite{simon2005} was for a different model than that considered here, the two models are quite close and the nature of this apparent discrepancy is not clear.  It is interesting to note that the authors of ref.~\onlinecite{simon2005} find very similar behaviour at $k_F = \pi / (2a)$ to that found by us for electron densities less than half-filled, $k_F < \pi / (2a)$ (see figure~\ref{fig:TKvsphi}).  It may be that the particle-hole symmetry breaking caused by moving away from half-filling in our calculation mimics the particle-hole symmetry breaking caused by the negative on-site energy of the additional site in the ring used in ref.~\onlinecite{simon2005}.  Perhaps it is this type of particle-hole symmetry breaking that leads to a flux dependent Kondo temperature.  This is speculation and further analysis of both methods would be required to resolve this apparent discrepency. 

Reference~\onlinecite{yoshii2008} follows a very similar procedure to that used here, transforming to the scattering basis and identifying the screening channel.  However, they mainly consider the $U \to \infty$ limit with finite dot energy level $\eps_d$.   Their subsequent scaling analysis, assuming half-filled leads with particle-hole symmetric Fermi energy $\eps_F = 0$, produces a flux dependent Kondo temperature.  Although this seems to contradict our conclusion that the Kondo temperature is flux \emph{independent} at half-filling, our result was obtained in a much different limit, with $\eps_d \approx -U/2$.  The authors do claim that, for finite $U$, the flux dependence is suppressed (though still present) when $\eps_d = -U/2$.  However, we find no evidence of any flux dependence in the Kondo temperature when $k_F = \pi / (2a)$.  

We close our discussion with a few comments on the apparent discrepancies presented in the NRG evidence for the additional potential scattering $V_R$.  As discussed in the text, we expect there to be cubic and higher order contributions to $V_R$ that we do not calculate so discrepencies for values of $\nu J$ that approach unity should be expected.  However, discrepancies remain even for relatively small values of $\nu J$ and we offer here some possibilities for why this might be.

As written at the end of section~\ref{NRGVR}, the correspondence between the tight-binding model used to derive $V_R$ in eq.~(\ref{VR}) and that used in the NRG is only approximate, especially for values of $\Lambda > 1$.  This leads to artificial $\Lambda$ dependence in many of the quantities extracted from the NRG as has been presented above.  This is probably true for the value of $V_R$ extracted from the NRG, as seen in figure~\ref{fig:VR_LambdaDependence}, suggesting that the form of $V_R$ may be non-universal in that it may depend on the details of the band structure of the leads.

To explore the universality of the form of $V_R$, the authors have repeated the derivation of $V_R$ for a model with a linear dispersion in the leads rather than the tight-binding cosine dispersion presented in the text.  It was found that, while qualitatively the same as the form of $V_R$ in eq.~(\ref{VR}), the two forms of $V_R$ did differ in numerical details.  From this we conclude that the form of $V_R$ is non-universal.  In light of this fact, one would ideally repeat the calculation of $V_R$, not for the tight-binding chain presented but for the full $\Lambda$-dependent Wilson chain and so obtain the $\Lambda$ dependence of $V_R$.  However, given the non-uniform `tunneling amplitudes' in the Wilson chain that go as $\Lambda^{-n/2}$ for hopping from the $n^{\mathrm{th}}$ site, such a calculation would be very difficult.  

Another possible source for this discrepancy is the possibility of additional contributions to potential scattering arising from the Schrieffer-Wolff transformation.  We have performed such a transformation to second order in $\tilde{V}_{dk}$ and concluded that the potential scattering $K_{k_F k_F}$ that arises vanishes when $\eps_d = -U/2$ so that, in this regime, $V_R$ contributes to the leading order term in the potential scattering.  However, given the fact that a non-zero $t^\p$ breaks particle-hole symmetry, there is nothing preventing the Schrieffer-Wolff transformation from generating a potential scattering term that is fourth order in $\tilde{V}_{dk}$ (equivalently, second order in $J$).  It would be interesting though non-trivial to carry out the Schrieffer-Wolff transformation to higher orders to see if indeed such potential scattering terms are present and if they can account for the disagreement with the NRG.  

Despite all of these possibilities, it is clear that such a $V_R$ term is present in both the tight-binding model as well as in the NRG and that they share the same qualitative behaviour and modestly agree quantitatively.  Given this, we expect such a $V_R$ term to be present in any real physical system and we expect it to share the same qualitative dependence on flux $\vphi$, inter-lead tunneling $t^\p$ (peaking around $t^\p \approx t$) and on electron density in the leads via the dependence on $k_F$ but do not claim that it will be precisely as that given in eq.~(\ref{VR}) which is based on an overly simplified tight-binding model.  Furthermore, although present, the contribution of $V_R$ to the conductance is very small for typical values of $\nu J$, as seen in figure~\ref{fig:Gflux}, and so will probably be difficult to detect explicitly in any physical system.  Nevertheless, the remainder of our analysis is robust and confirmed numerically and provides a framework in which to think about such quantum dot systems. 

\section*{Acknowledgements}
The authors would like to thank Josh Folk, Eran Sela, and Jan von Delft for fruitful discussions.  This work was supported by the Natural Sciences and Engineering Research Council of Canada (JM and IA), the Canadian Institute for Advanced Research (IA), and the Government of British Columbia (JM).

\appendix

\section{Details of the transformation to the scattering basis}
\label{sec:tmatrixderive}
Let us consider the Hamiltonian $H = H_0 + H_{-+}$ where $H_0$ and $H_{-+}$ are given by eq.~(\ref{H0eo}) and~(\ref{H+-ab}) respectively.   We will now demonstrate that, under the transformation of eqs.~(\ref{qak})--(\ref{qbk}), the above Hamiltonian takes the form
\be
H = \int_0^{\frac{\pi}{a}} dk \, \eps_k \left(q_{ak}^\dag q_{ak} + q_{bk}^\dag q_{bk} \right).
\ee
Another way of saying this is that we demand the transformation to be such that
\be
\label{comq}
\com{H}{q_{\alpha k}^\dag} = \eps_k q_{\alpha k}^\dag \qquad , \alpha = a, b.
\ee

Substituting into eq.~(\ref{comq}) the definition of $q_{\alpha k}^\dag$ in terms of $e_k^\dag$ and $o_k^\dag$ from eqs.~(\ref{qak})--(\ref{qbk}) and using the relations
\bea
\com{H}{e_k^\dag} & = & \eps_k e_k^\dag + \int_0^{\frac{\pi}{a}} dk^\p v_{k^\p k} e_{k^\p}^\dag \\
\com{H}{o_k^\dag} & = &  \eps_k o_k^\dag - \int_0^{\frac{\pi}{a}} dk^\p v_{k^\p k} o_{k^\p}^\dag
\eea
one obtains
\be
\left( \eps_k - \eps_{k^\p} \right) \phi_{k^\p}^{\pm (k)} = \pm \int_0^{\frac{\pi}{a}} dq \, \phi_q^{\pm (k)} v_{k^\p q}.
\ee
Substituting now the definition of $\phi_{k^\p}^{\pm (k)}$ from eq.~(\ref{phikkp}) in the above expression gives an integral equation for $T_{k k^\p}^\pm$
\be
\label{Tintegraleq}
T^\pm_{k k^\p} = \pm \left( v_{k k^\p} + \int_0^{\frac{\pi}{a}} dq \frac{ v_{k q} T^\pm_{q k^\p} }{ \eps_{k^\p} - \eps_q + i \eta } \right).
\ee

If we now take the ansatz
\be
\label{Tansatz}
T_{k k^\p}^\pm = T^\pm_{k^\p} \sin ka
\ee
and substitute this into eq.~(\ref{Tintegraleq}) together with the definition of $v_{k k^\p}$ from eq.~(\ref{vkkp}) and $\eps_k = - 2 t \cos ka$, one obtains the following equation for $T_{k}$
\be
\label{Tkstart}
T^\pm_k = \mp \frac{2 t^\p a}{\pi} \frac{ \sin ka}{1 \pm \frac{\tau^\p}{\pi}I_k}
\ee
where, as before, $\tau^\p \equiv t^\p / t$ and $I_k$ is the dimensionless integral 
\be
\label{Ikdef}
I_k \equiv \frac{1}{2} \int_{-\pi}^\pi dy \frac{ \sin^2 y}{\cos y - \cos k a + i \eta }
\ee 
This integral can be solved in the complex plane.  Making the change of variables $z = e^{iy}$, one can write this as
\be
I_k = - \frac{1}{4i} \oint dz \frac{ \left( z^2 - 1 \right)^2 }{ z^2 \left( z^2 - 2z \left( \cos ka - i \eta \right) + 1 \right)}
\ee
where the contour of integration is the unit circle centered at the origin in the complex $z$ plane.  

The integrand has poles at $z = z_0 = 0$ (second order) and at $z = z_\pm$ (simple), the latter given by
\be
z_\pm = \cos ka \pm i | \sin ka | - i \eta \mp \eta \frac{\cos ka}{| \sin ka |}.
\ee
Since $\eta$ is a positive infinitesimal quantity, one can show that $|z_+| < 1$ whereas $|z_-| > 1$ so that only the $z_+$ and $z_0$ poles lie within the contour.  Applying the residue theorem
\bea
I_k & = & - \frac{\pi}{2} \left( \mathrm{Res} ( z = z_0 ) + \mathrm{Res} ( z = z_+ ) \right) \\ 
\label{Ikvalue}
& = & \pi e^{-i ka}. 
\eea
Substituting this final value back into eq.~(\ref{Tkstart}) and the resulting $T_k$ back into the ansatz, eq.~(\ref{Tansatz}), produces the promised form of $T_{k k^\p}^\pm$
\be
\label{Tapp}
T^\pm_{k k^\p} = \mp \frac{2 t^\p a}{\pi} \frac{ \sin ka \sin k^\p a}{1 \pm \tau^\p e^{-i k^\p a} }
\ee
as stated in eq.~(\ref{Tkkp}).

Finally, we compute the functional form of $\Gamma_k^\pm$ as defined and stated in eq.~(\ref{Gammakdef}).  Substituting the definition of $\phi_{k^\p}^{\pm (k)}$ as defined in eq.~(\ref{phikkp}) into the definition of $\Gamma_k^\pm$ and using the derived form of $T^\pm_{k k^\p}$, eq.~(\ref{Tapp}),  gives
\bea
\Gamma_k^\pm & \equiv & \int_0^{\frac{\pi}{a}} dk^\p \, \sin k^\p a \, \phi_{k^\p}^{\pm (k)} \\
& = & \sin ka \left(1 \mp \frac{\tau^\p}{\pi} \frac{I_k}{1 \pm \tau^\p e^{-i ka }} \right)
\eea
where $I_k$ is the same dimensionless integral defined in eq.~(\ref{Ikdef}) and computed in eq.~(\ref{Ikvalue}).  Hence, the final result is obtained upon substitution
\be
\Gamma_k^\pm = \frac{ \sin ka }{ 1 \pm \tau^\p e^{-i ka} }
\ee
as reported in eq.~(\ref{Gammakdef}).

\section{Potential scattering phase shift}
\label{sec:phaseshift}

\subsection{Lattice Model}
Consider a single semi-infinite tight-binding chain with an on-site potential at the first site
\be
H = -t \sum_{j=1}^\infty \left( e_j^\dag e_{j+1} + \hc \right) - t^\p e_1^\dag e_1.
\ee
This is the same as the even-channel Hamiltonian $H = H_0 + H_{-+}$ of eqs.~(\ref{H0eo})--(\ref{H+-ab}) in the limit $t_{d+} = t_{d-} = 0$.  The presence of a finite $t^\p$ will give rise to a phase shift in the single-particle wave function and it is the calculation of this phase shift that is the subject of this appendix.

We write the eigenvectors of the Hamiltonian as
\be
\ket{\phi} = \sum_{j=1}^\infty \phi_j e_j^\dag \ket{0}
\ee
where $\{ \phi_j \}$ are coefficients to be determined such that they satisfy the Schr\"odinger equation
\be
H \ket{\phi} = \eps_k \ket{\phi}
\ee
with $\eps_k = - 2 t \cos ka$.  The Schr\"odinger equation can be written as the following series of algebraic equations
\bea
\label{deltaeq1}
-t \phi_2 - t^\p \phi_1 & = & \eps_k \phi_1 \\
-t \left( \phi_{j-1} + \phi_{j+1} \right) & = & \eps_k \phi_j \qquad, j > 1.
\eea
The equations in the second line can be solved by taking coefficients of the form 
\be
\phi_j = \sin \left( kja + \delta_k \right)
\ee
and $\delta_k$ is determined by eq.~(\ref{deltaeq1}) to be
\be
\tan \delta_k = \frac{ \tau^\p \sin ka}{1 - \tau^\p \cos ka}
\ee
where, as before, $\tau^\p = t^\p / t$.  This is the form of the phase shift $\delta^+_k$ that occurs in the even channel.  The phase shift in the odd channel is the same but with $\tau^\p \to - \tau^\p$ so that
\be
\tan \delta^\pm_k = \pm \frac{ \tau^\p \sin ka}{1 \mp \tau^\p \cos ka}.
\ee

\subsection{Continuum Model}
At low energies (long wavelengths), one can take the continuum limit of the tight-binding model and linearize the dispersion relation about $k_F$.  In this way, one can write an approximate real-space Hamiltonian
\be
H = v_F \int_{-\infty}^\infty dx \, \psi^\dag(x) (-i \partial_x ) \psi(x) + V_R \psi^\dag(0) \psi(0)
\ee
where $v_F$ is the Fermi velocity.   We assume that $\nu V_R \ll 1$ where $\nu$ is the density of states at the Fermi energy. 

As in the lattice model, we introduce eigenvectors of the Hamiltonian
\be
\ket{\phi_k} = \int_{-\infty}^\infty dx \phi_k (x) \psi^\dag(x) \ket{0}
\ee
which satisfy the Schr\"odinger equation
\be
H \ket{\phi_k} = v_F k \ket{k}.
\ee
This puts the following condition on the functions $\phi_k(x)$
\be
\label{condition}
-i v_F \partial_x \phi_k(x) + V_R \phi_k(0) \delta(x) = v_F k \phi_k(x).
\ee

We now take the ansatz
\be
\phi_k(x) = \left\{ 
\begin{array}{ll}
e^{i(kx + \delta_R)} & \quad , x>0 \\
e^{i(kx - \delta_R)} & \quad , x<0 \\
\cos \delta_R & \quad , x=0
\end{array}
\right.
\ee
with derivative
\be
\partial_x \phi_{k}(x) = \left\{
\begin{array}{ll}
ik e^{i(kx + \delta_R)} & \quad , x>0 \\
ik e^{i(kx - \delta_R)} & \quad , x<0 \\
2i \sin \delta_R \delta(x) & \quad , x=0.
\end{array}
\right.
\ee
In order for $\phi_k(x)$ to satisfy eq.~(\ref{condition}), we require
\be
\tan \delta_R = - \frac{V_R}{2 v_F} = - \pi \nu V_R.
\ee


\begin{thebibliography}{99}
\bibitem{footnote1} The odd $N$ phase shifts are related by a shift of $\pi / 2$.
\bibitem{footnote2} However, there may be other contributions to the potential scattering that are of the same order as $V_R$ as discussed in \S~\ref{sec:discussion}.
\end{thebibliography}
\end{document}